\documentclass{article} 
\usepackage{a4wide}
\usepackage{mncite}
\usepackage{epsf}
\usepackage{amsmath}
\usepackage{amssymb}

\def\hMpc{\ifmmode{h^{-1}{\rm Mpc}}\else{$h^{-1}$Mpc}\fi}
\def\e{{\rm e}}
\def\cA{{\mathcal{A}}}
\def\cB{{\mathcal{B}}}
\def\deg{^{\rm o}}

\begin{document}

\title{The Significance of the Fluctuations\\
in the IRAS 1.2 Jy Galaxy Catalogue\footnote{Proc.\ $2^{\rm nd}$ SFB
workshop on {\em Astro--particle physics} Ringberg 1996, Report SFB
375/P002 (1997), R.~Bender, T.~Buchert, P.~Schneider (eds.), in
press.} }

\author{Martin Kerscher$^1$, Jens Schmalzing$^{1,2}$, 
Thomas Buchert$^1$ \& Herbert Wagner$^1$\\
\small
$^1$Ludwig--Maximilians--Universit\"at, \\
\small
Theresienstr.~37, 80333 M\"unchen, Germany\\[1ex]
\small
$^2$Max--Planck--Institut f\"ur Astrophysik,\\
\small
Karl--Schwarzschild--Stra{\ss}e 1, 85740 Garching, Germany\\
\small
emails: kerscher, jens, buchert, wagner@stat.physik.uni-muenchen.de}

\maketitle

\begin{abstract}
In an analysis of the IRAS 1.2~Jy redshift catalogue, with emphasis on
the separate examination of northern and southern parts (in galactic
coordinates), we found that the clustering of galaxies differs
significantly between north and south, showing fluctuations in the
clustering properties at least on scales of 100\hMpc\
{}\cite{kerscher:fluctuations}. In Section~\ref{sect:def_mink} we give
a brief description of our morphological method which is based on
Minkowski functionals; in Section~\ref{sect:result_100} we show the
results obtained from the IRAS 1.2~Jy galaxy catalogue.
Section~\ref{sect:error} contains a discussion of several error
estimates and in Section~\ref{sect:selection} we select different
subsamples from the 1.2~Jy catalogue according to flux and colour and
validate the results of Section~\ref{sect:result_100}. Furthermore we
look closer at the spatial origin of the fluctuation in
Section~\ref{sect:origin} and we compare with optically selected
galaxies from the CfA1 survey in Section~\ref{sect:jy-cfa}.
\end{abstract}

\section{Minkowski functionals}
\label{sect:def_mink}
We consider a set $\{{\mathbf x}_i\}_{i=1}^{N}$ of $N$ points
given by the redshift--space coordinates of the galaxies in the IRAS
1.2 Jy catalogue {}\cite{fisher:irasdata}. To characterize the
properties of this point set with the help of Minkowski functionals
introduced into cosmology by {}\scite{mecke:robust}, we decorate each
point ${\mathbf x}_i$ with a ball $\cB_r({\mathbf x}_i)$ of radius~$r$
and consider the union set $\cA_N(r) = {}\bigcup_{i =0}^N
\cB_r({\mathbf x}_i)$. {}\scite{hadwiger:vorlesung} proved
that in three--dimensional space the four Minkowski functionals
$M_{\mu=0,1,2,3}(\cA_N(r))$ give a complete morphological
characterization of the body $\cA_N(r)$. For the interpretation of
these functionals in terms of geometrical and topological quantities
see {}\scite{mecke:robust} and Table
\ref{table:mingeom}.
\begin{table}
\begin{center}
\begin{tabular}{|cl|c|c|c|}
\hline
       & geometric quantity      & $\mu$ & $M_\mu$      & $\Phi_\mu$   \\[1ex]
\hline
$V$    & volume                  & 0 	 & $V$ & $V/(\tfrac{4\pi}{3}r^3N)$ \\
$A$    & surface                 & 1 	 & $A/8$        & $A/(4\pi r^2N)$  \\
$H$    & integral mean curvature & 2 	 & $H/(2\pi^2)$   & $H/(4\pi rN)$    \\
$\chi$ & Euler characteristic    & 3 	 & $3\chi/(4\pi)$ & $\chi/N$       \\
\hline
\end{tabular}
\end{center}
\caption{\label{table:mingeom}
Minkowski functionals in three--dimensional space expressed in terms
of more familiar geometric quantities.}
\end{table}
Reduced, dimensionless Minkowski functionals $\Phi_\mu(\cA_N(r))$ may
be constructed by normalizing with the Minkowski functionals $M_\mu(\cB_r)$
of balls,
\begin{equation}\label{eq:Phi-def}
 \Phi_\mu(\cA_N(r)) := \frac{M_\mu(\cA_N(r))}{N M_\mu(\cB_r)}.
\end{equation}
For a Poisson process the functionals can be calculated analytically
(see {}\pcite{mecke:euler}) with the results:
\begin{equation}\label{eq:Poisson}
\begin{array}{rclrcl}
\Phi_0^{\rm P} & = & \left(1 - \e^{-\eta}\right)\ \eta^{-1}, &
\Phi_1^{\rm P} & = & \e^{-\eta} ,\\[1ex]
\Phi_2^{\rm P} & = & \e^{-\eta}\ (1 - \frac{3 \pi^2}{32} \eta ), &
\Phi_3^{\rm P} & = & \e^{-\eta}\ (1 - 3 \eta + \frac{3 \pi^2}{32} \eta^2 ), 
\end{array}
\end{equation}
where $\eta := \overline n M_0(\cB_r) = \overline n \ 4 \pi r^3/3$,
with the mean number density $\overline n$. As seen from
Eqs.~(\ref{eq:Poisson}), the Minkowski functionals
$\Phi_\mu(\cA_N(r))$, $\mu = 1,2,3$, which are supported by the
surface, are proportional to $\e^{-\eta(r)}$ for a Poisson process. A
similar exponential decay is also found in the case of more general
cluster processes. Since we are interested in the behavior of the
point process on large--scales we remove the exponential term
$\e^{-\eta}$ and consider functionals $\phi_{\mu=1,2,3}$, defined by
\begin{equation}\label{eq:phi-def}
 \phi_\mu(\cA_N(r)) = \frac{\Phi_\mu(\cA_N(r))}{\Phi_1^{\rm P}(r)}.
\end{equation}

\section{Results from the 100\hMpc\ sample}
\label{sect:result_100}
We consider a volume limited sample of the IRAS 1.2 Jy galaxy
catalogue \cite{fisher:irasdata} with limiting depth of 100\hMpc\
(H$_0$=100\ $h$\ km\ s$^{-1}$ Mpc$^{-1}$).

\begin{figure}
 \begin{center} 
 \epsfxsize=7.4cm
 \begin{minipage}{\epsfxsize}\epsffile{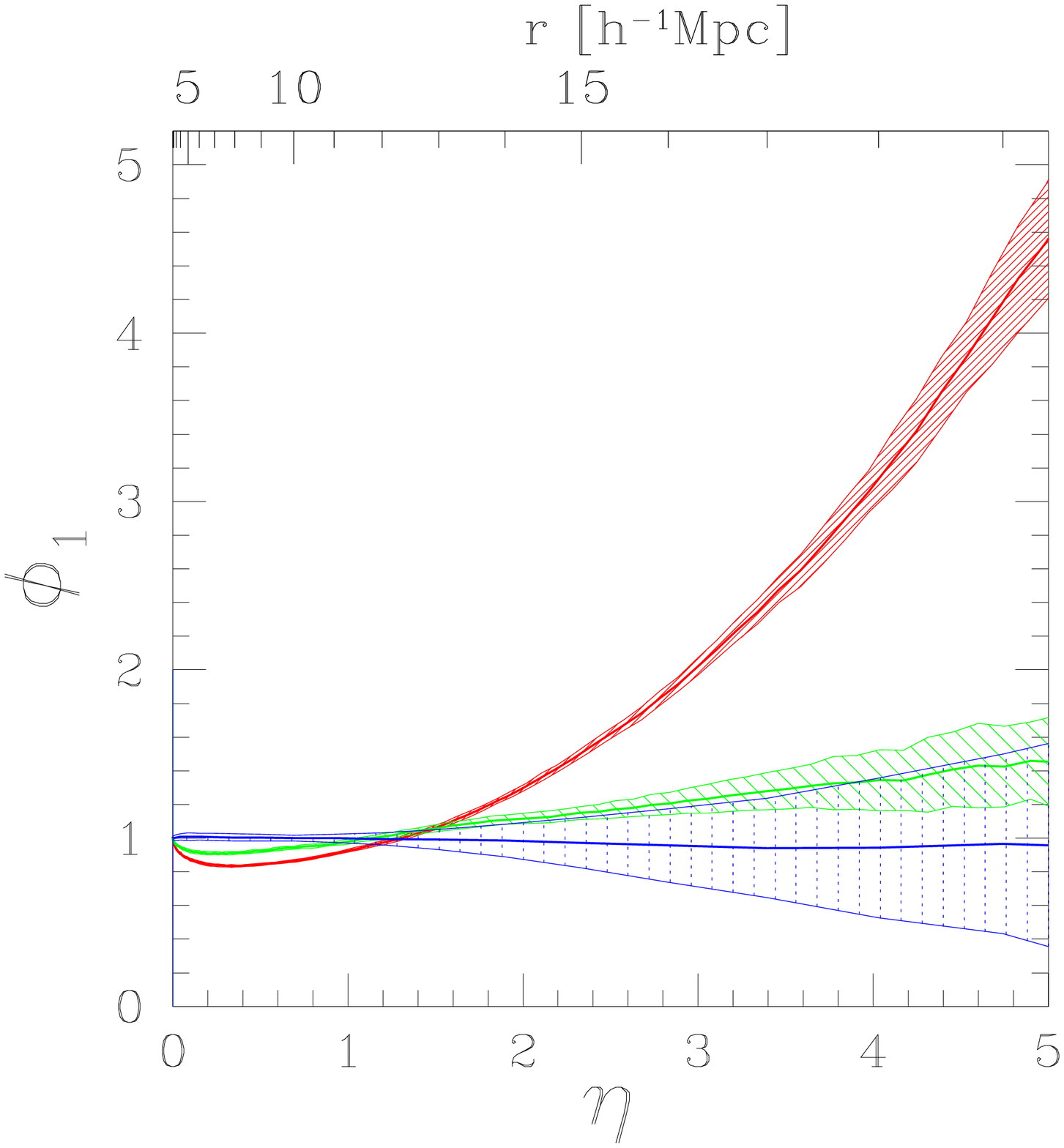}\end{minipage} 
 \epsfxsize=7.4cm
 \begin{minipage}{\epsfxsize}\epsffile{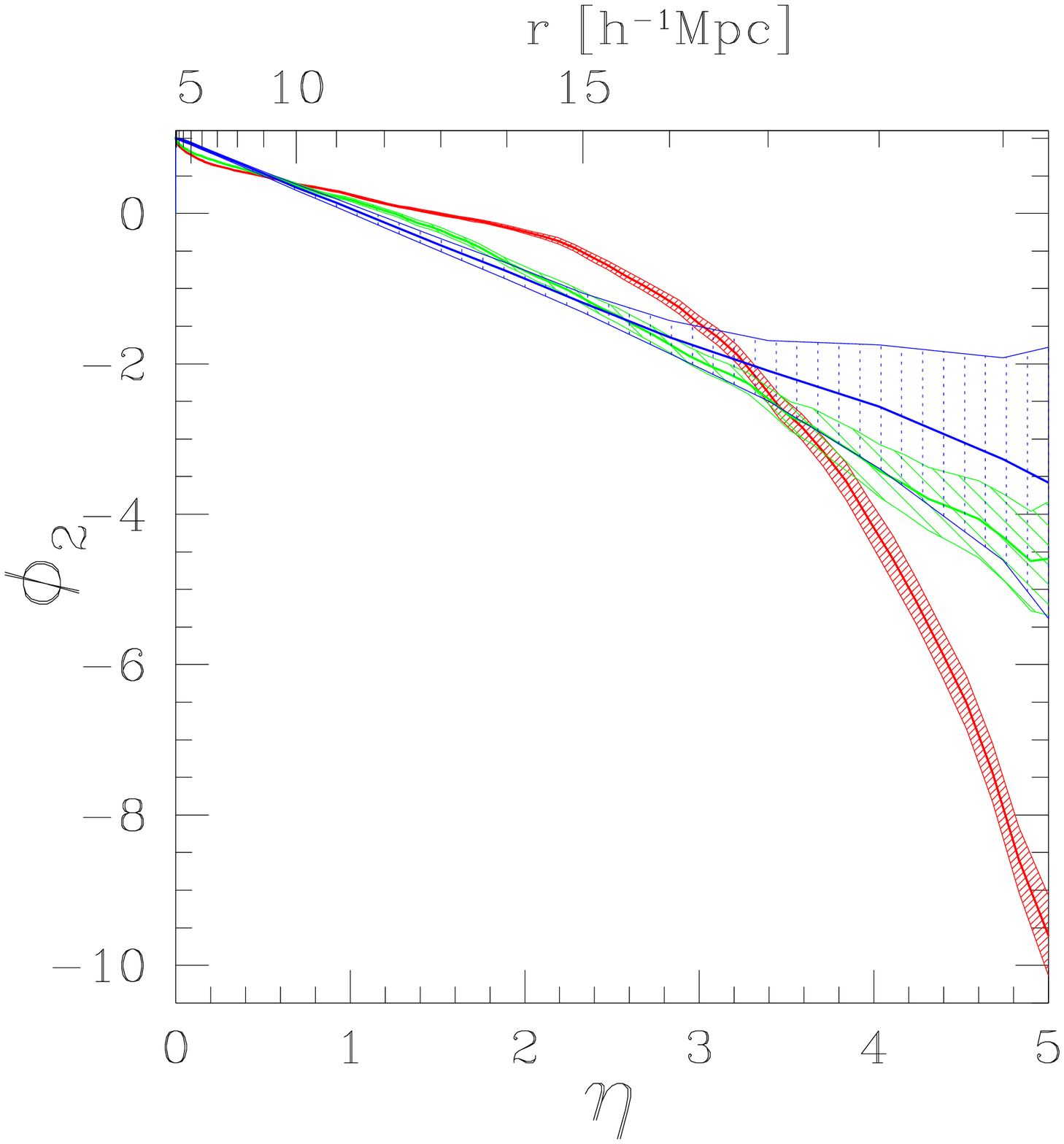}\end{minipage} 
 \epsfxsize=7.4cm
 \begin{minipage}{\epsfxsize}\epsffile{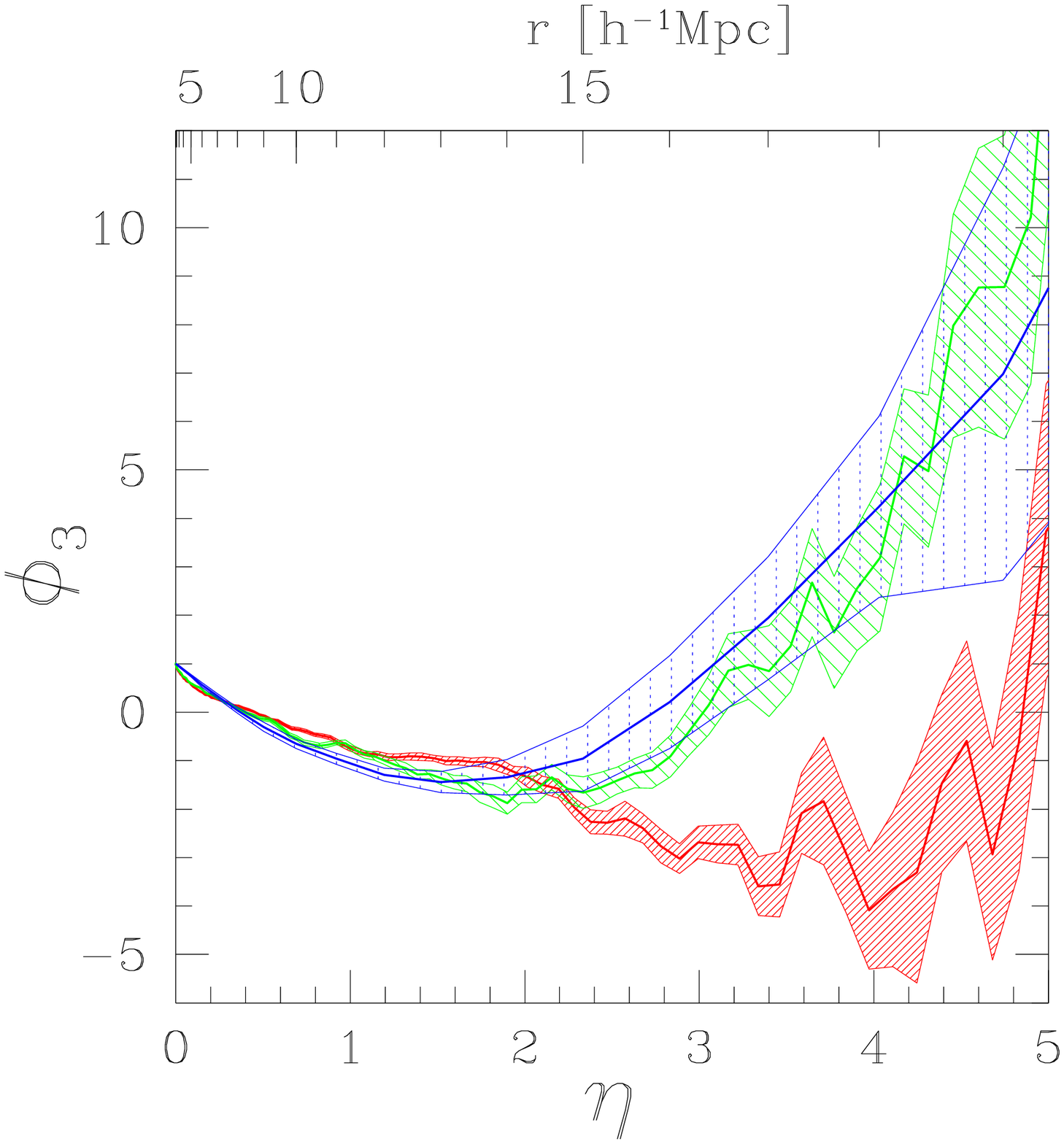}\end{minipage} 
 \end{center}
\caption{\label{fig:minjyv10}
Minkowski functionals $\phi_\mu$ of a volume limited sample with
$100\hMpc$ depth; dark shaded areas: southern part; medium shaded
areas: northern part; dotted areas: Poisson process with the same
number density. The areas correspond to $1\sigma$ errors, as explained
in the text.}
\end{figure}

Fig.~\ref{fig:minjyv10} displays the values of the Minkowski
functionals of the southern and northern parts in comparison with the
functionals of a Poisson process with the same number density. The
northern part contains 352 galaxies, and the southern part 358
galaxies. The errors of the Poisson process were calculated from
twenty different realizations. To estimate the error from the
catalogue data we calculated the Minkowski functionals of twenty
subsamples containing 90\% of the galaxies, randomly chosen from the
volume limited subsample (see Section \ref{sect:error}).

In both parts of the 1.2~Jy catalogue the galaxy clustering on scales
up to 10\hMpc\ is distinctly stronger than in the case of a Poisson
process, as inferred from the lower values of the surface functional
$\phi_1$, the integral mean curvature functional $\phi_2$ and the
Euler characteristic functional $\phi_3$. Moreover, the northern and
southern parts differ significantly in their morphological features,
with the northern part being less clumpy. The most conspicuous
features are the enhanced surface area $\phi_1$ in the southern part
on scales from 12 to 20~\hMpc\ and the decrease of the integral mean
curvature $\phi_2$ which sets in at 14~\hMpc. This behaviour indicates
that dense substructures in the southern part fill up at this scale
(i.e.~the balls in these substructures overlap strongly); this is
probably the signature of the Perseus--Pisces supercluster (compare
Section~\ref{sect:origin}). On scales from 15 to 20~\hMpc, the
integral mean curvature $\phi_2$ is negative indicating concave
structures. In the southern part the Euler characteristic is negative
in this range, therefore, the structure is dominated by interconnected
tunnels (negative contributions to the Euler characteristic;
completely enclosed voids would yield positive contributions).

\section{Error estimates}
\label{sect:error}
\begin{figure}
 \begin{center} 
 \epsfxsize=7.4cm
 \begin{minipage}{\epsfxsize}\epsffile{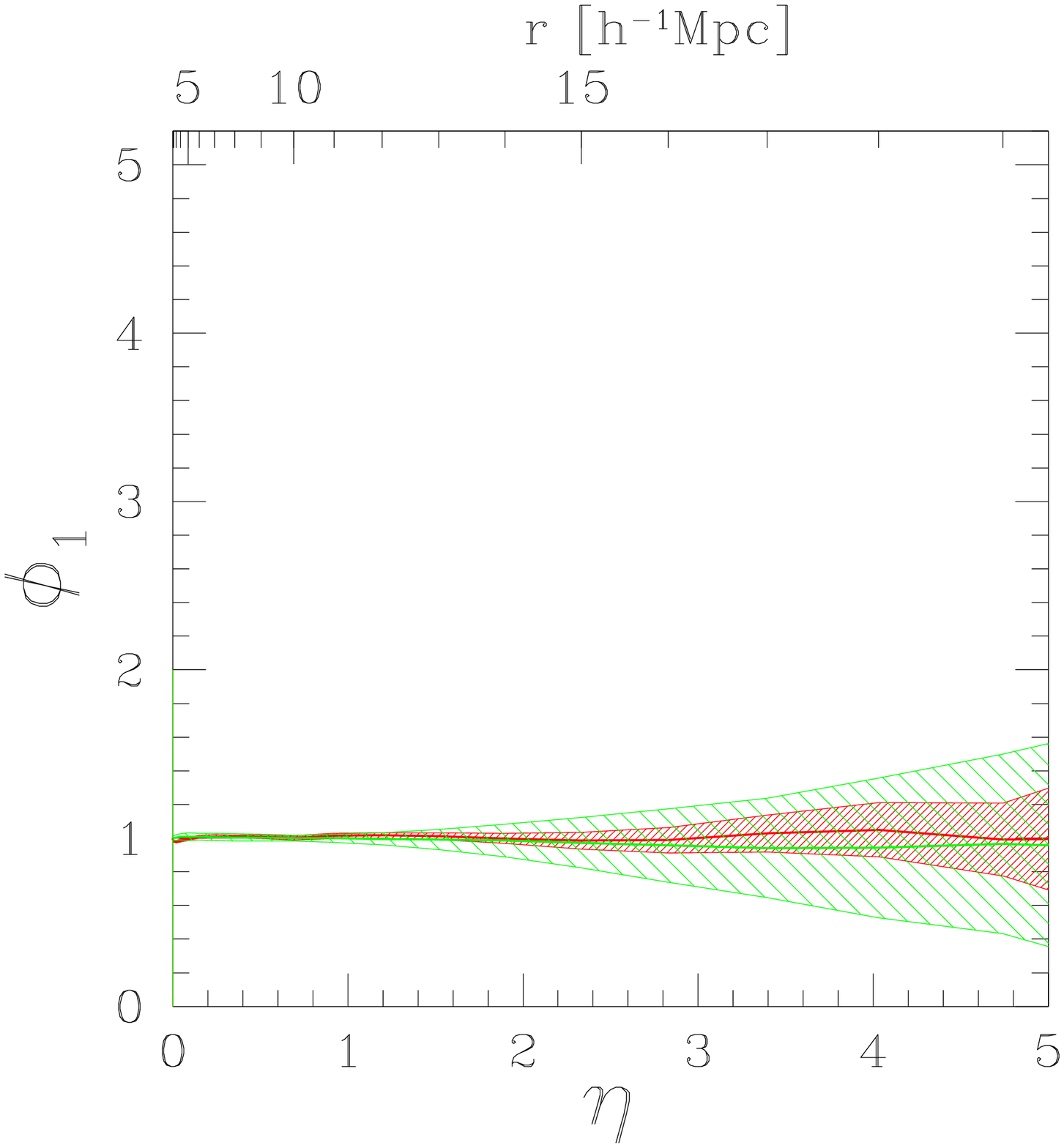}\end{minipage} 
 \epsfxsize=7.4cm
 \begin{minipage}{\epsfxsize}\epsffile{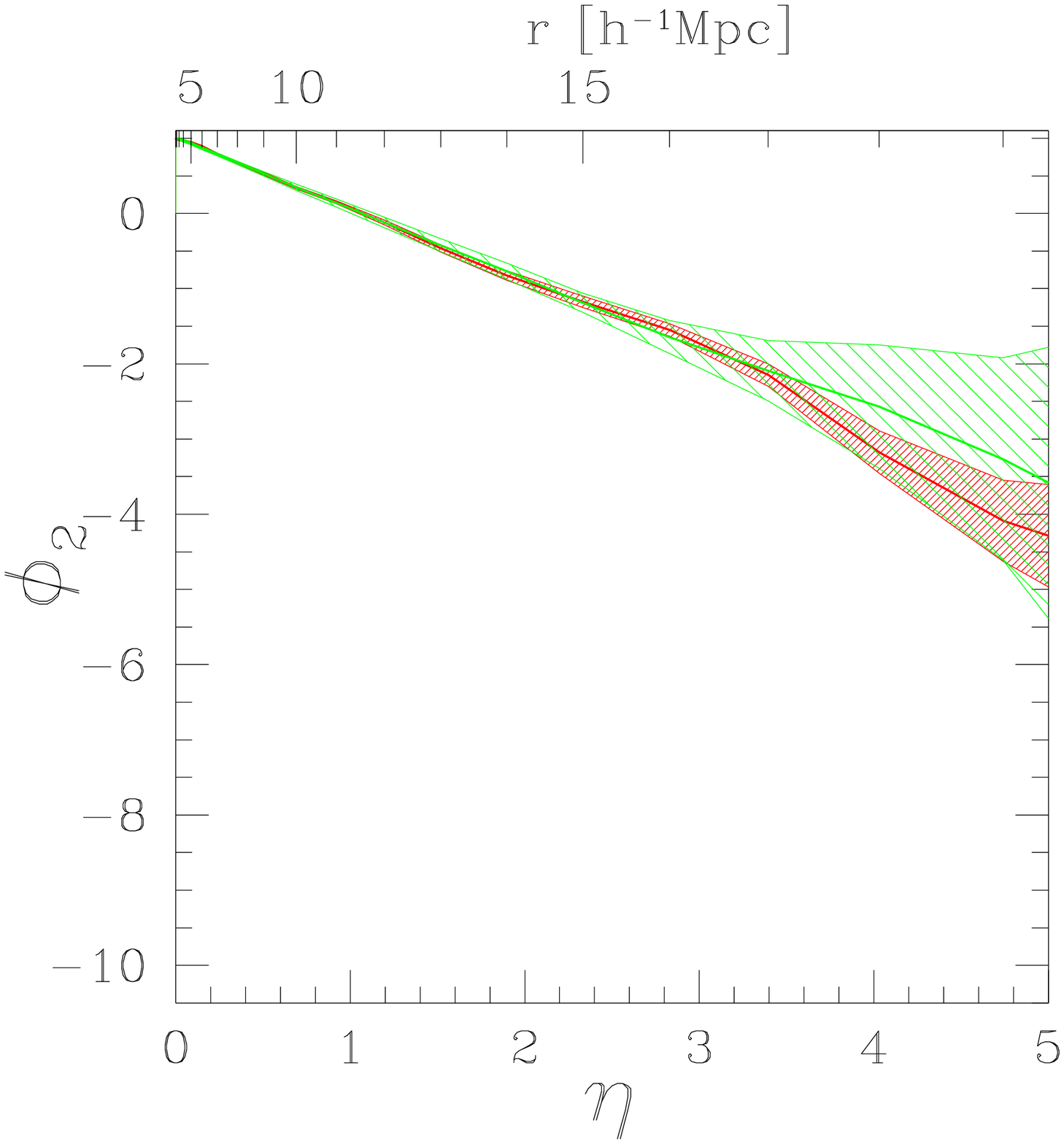}\end{minipage} 
 \epsfxsize=7.4cm
 \begin{minipage}{\epsfxsize}\epsffile{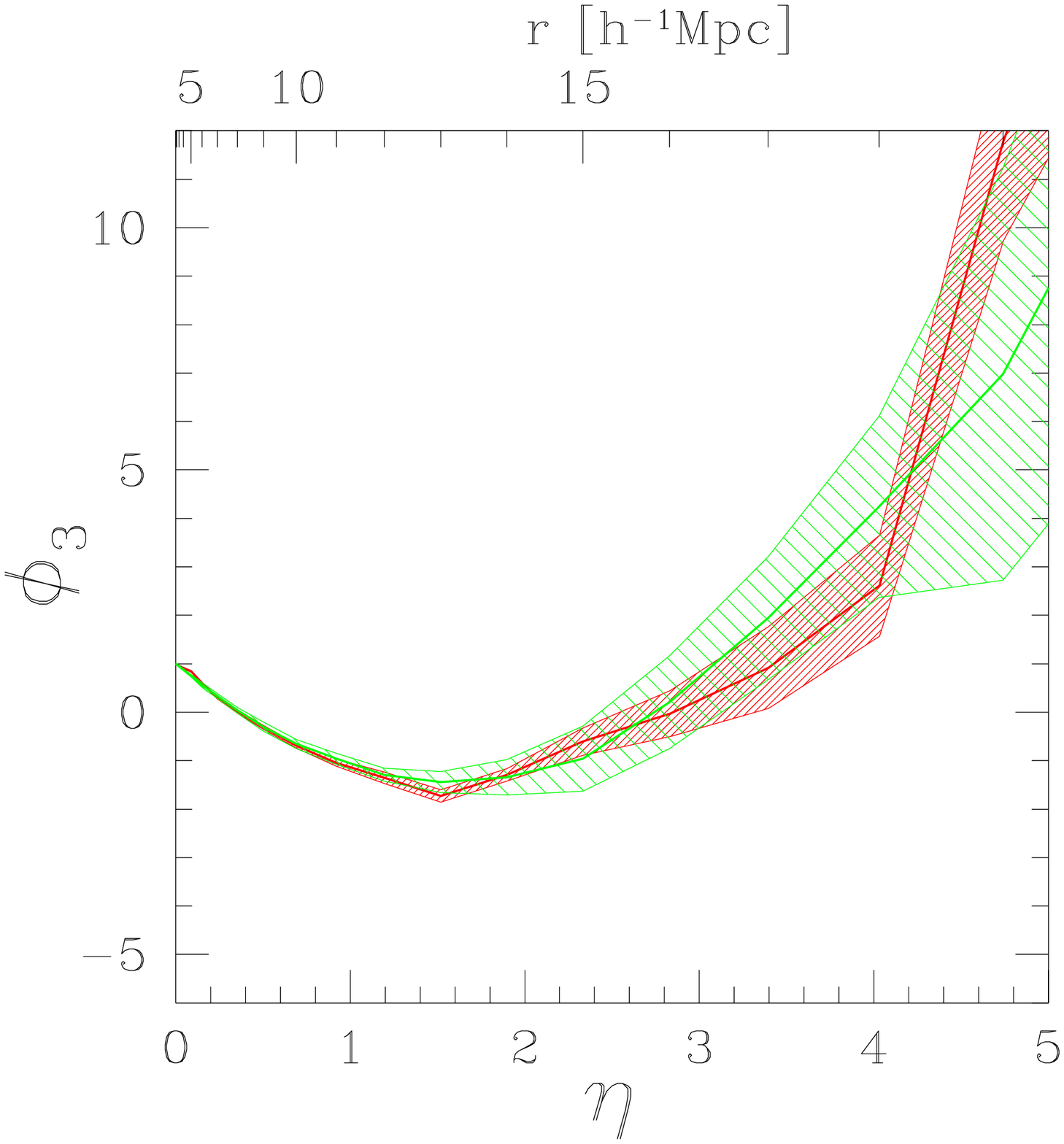}\end{minipage} 
 \end{center}
\caption{\label{fig:poi_sub}
Minkowski functionals $\phi_\mu$ of Poisson process realizations; the
dark shaded area is the $1\sigma$ error determined with 90\%
subsampling out of one realization, the light shaded area represents
the $1\sigma$ ensemble error estimated from 20 different realizations.}
\end{figure}

To find out how the error from subsampling is related to
the intrinsic variance of an ensemble we took fixed realizations of a
Poisson process within the sample geometry (with the same number of
galaxies as in the volume limited samples) and calculated the error
via subsampling, using again 90\% of the points (see
Figure~\ref{fig:poi_sub}). This error turns out to be about two times
smaller than the ensemble error estimated from 20 realizations of a
Poisson process with the same number density.
A comparison with the variance of a Poisson process may be helpful,
but the fluctuations of the functionals calculated for the galaxy
distribution exceed this variance (compare the fluctuations between
north and south with the variance of the Poisson process in
Fig.~\ref{fig:minjyv10}). With the errors calculated from subsampling
we only check whether our estimators are robust; the fluctuations may
be significantly larger.

\begin{figure}
 \begin{center} 
 \epsfxsize=7.4cm
 \begin{minipage}{\epsfxsize}\epsffile{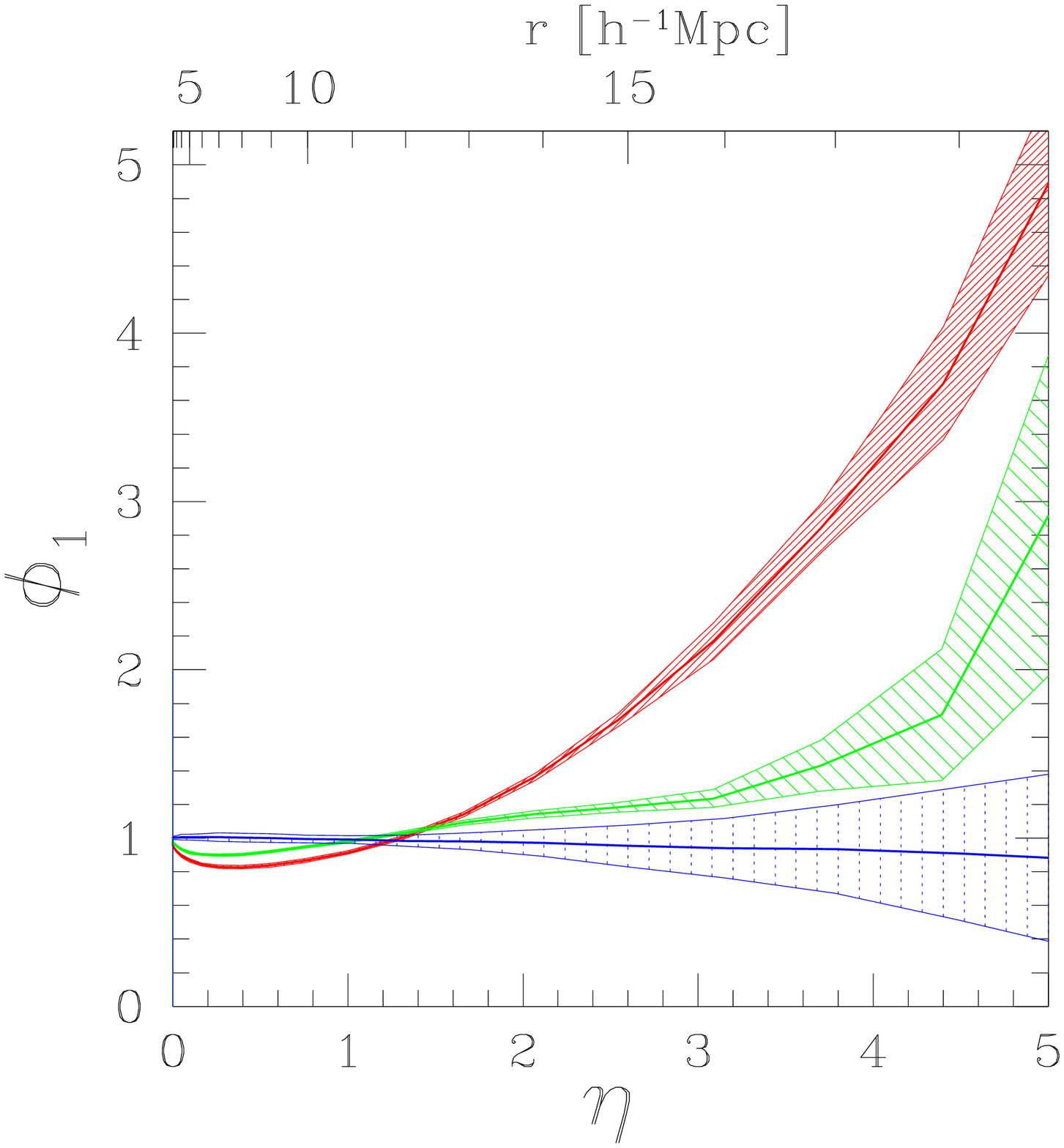}\end{minipage} 
 \epsfxsize=7.4cm
 \begin{minipage}{\epsfxsize}\epsffile{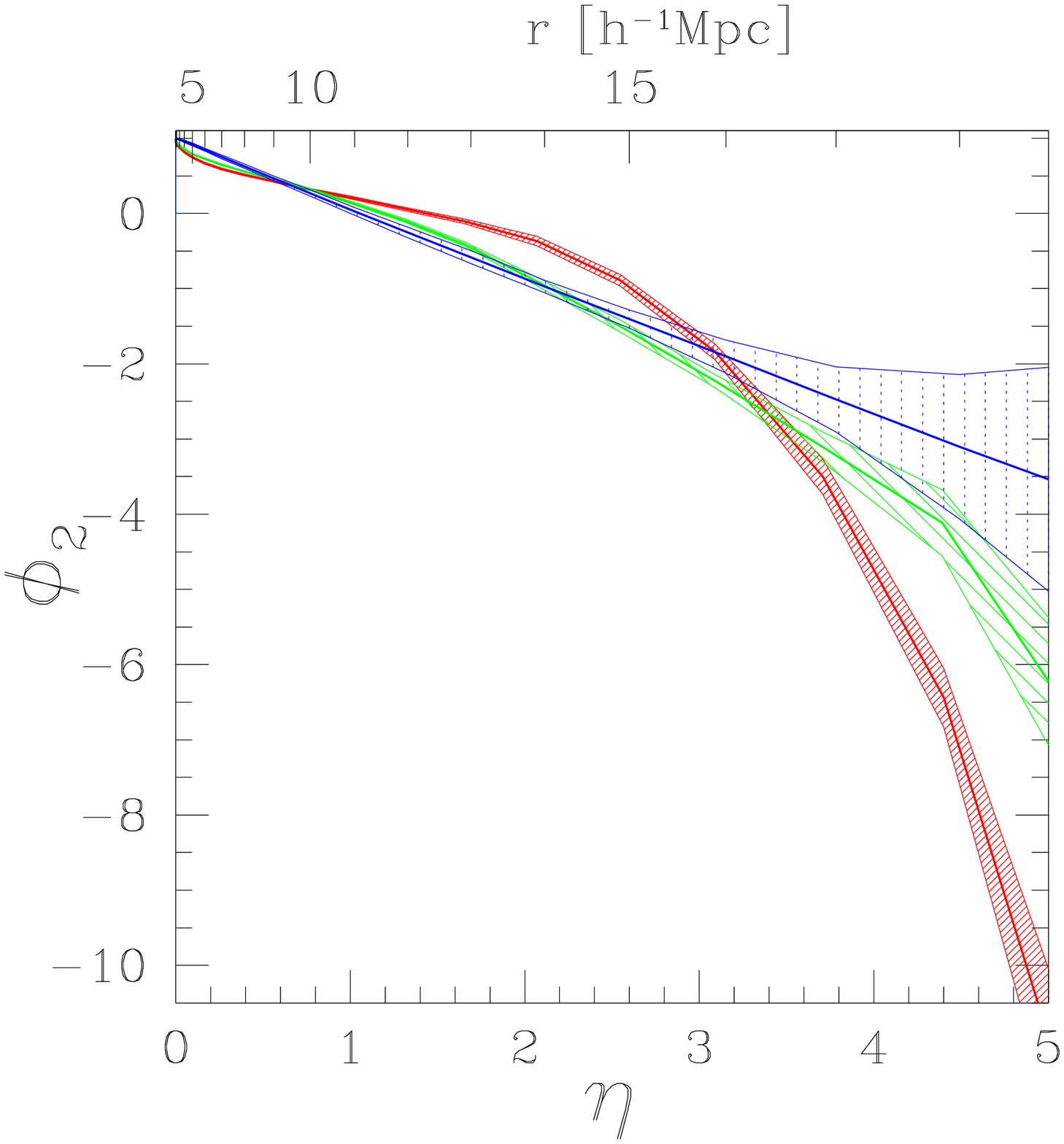}\end{minipage} 
 \epsfxsize=7.4cm
 \begin{minipage}{\epsfxsize}\epsffile{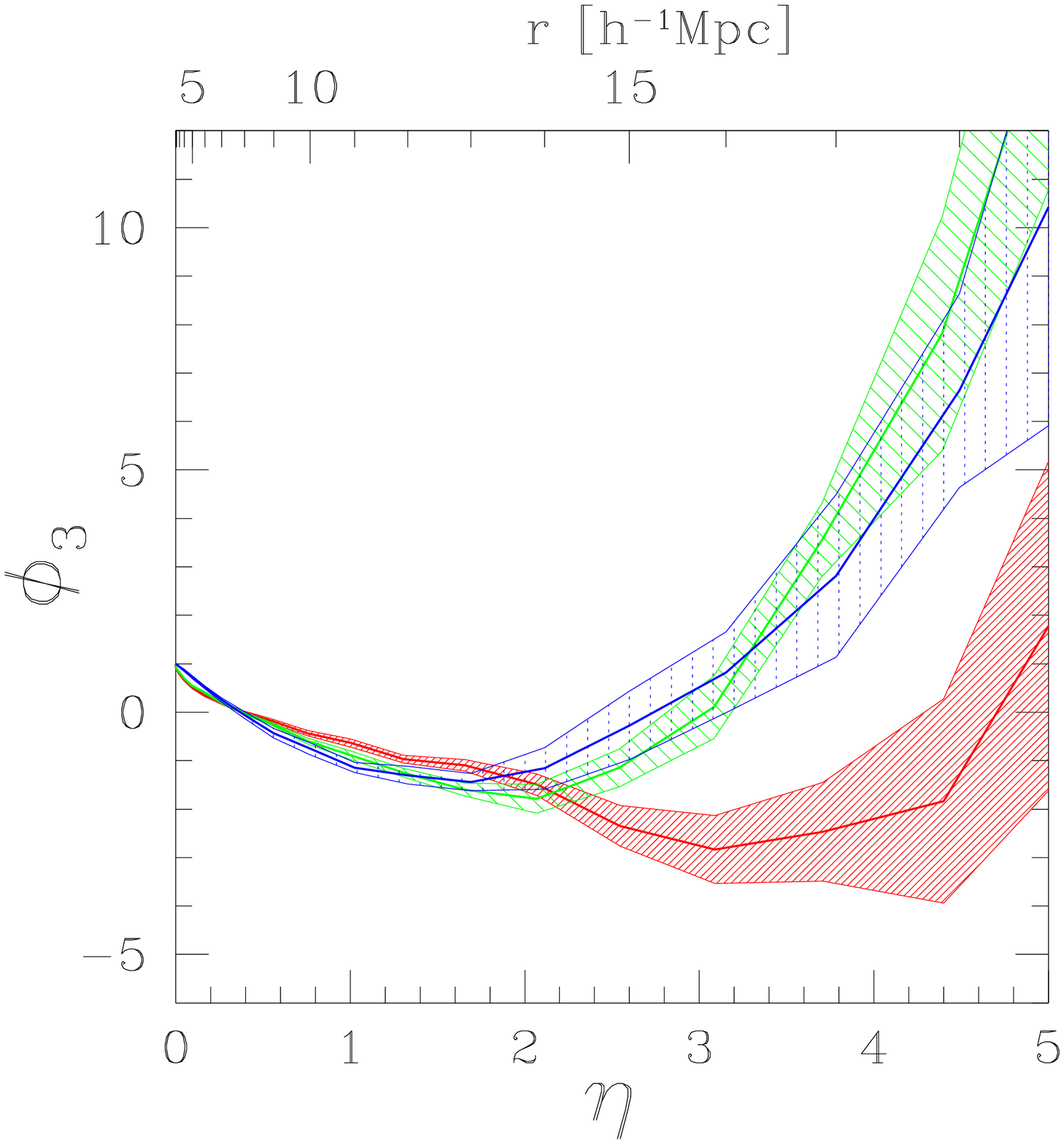}\end{minipage} 
 \end{center}
\caption{\label{fig:minjyv10E5}
Minkowski functionals $\phi_\mu$ of a volume limited sample with
$100\hMpc$ depth; now the error is calculated via randomizing the
redshifts with five times the quoted error. The shaded areas are the
$1\sigma$ errors as in Figure~\ref{fig:minjyv10}.}
\end{figure}

To obtain another estimate for measurement errors, we randomized the
redshifts using the quoted redshift errors as the standard deviation
(and using the mean redshift error, if none is quoted). These errors
are approximately two times smaller than the errors from
subsampling. Even if we increase the quoted redshift error by a factor
of five, the errors in the values of the functionals are only of the
same order as determined from subsampling (compare
Fig.~\ref{fig:minjyv10} and Fig.~\ref{fig:minjyv10E5}).

We also took care for incomplete sky coverage. Apart from a 10 degree
wide zone of avoidance, whose boundary effects are completely removed
according to {}\scite{schmalzing:minkowski}, there are additional
empty regions in the redshift catalogue due to a lack of sky coverage
or confusion in the point source catalogue. In the northern part these
regions account for 3.2~\% of the sky in the southern part they
account for 4.5~\%. To estimate the influence of these regions on the
statistics we added Poisson distributed points with the same number
density. The additional error introduced from these random points is
much smaller than the errors from subsampling. Moreover, no systematic
effect is seen, the curves overlap completely.

\section{Selection effects}
\label{sect:selection}
Several selection effects might enter into the construction of the
catalogue. Therefore, we draw subsamples selected according to special
features like ``colours'' or higher limiting flux.

\begin{figure}
\begin{center} 
\epsfxsize=7.4cm
\begin{minipage}{\epsfxsize}\epsffile{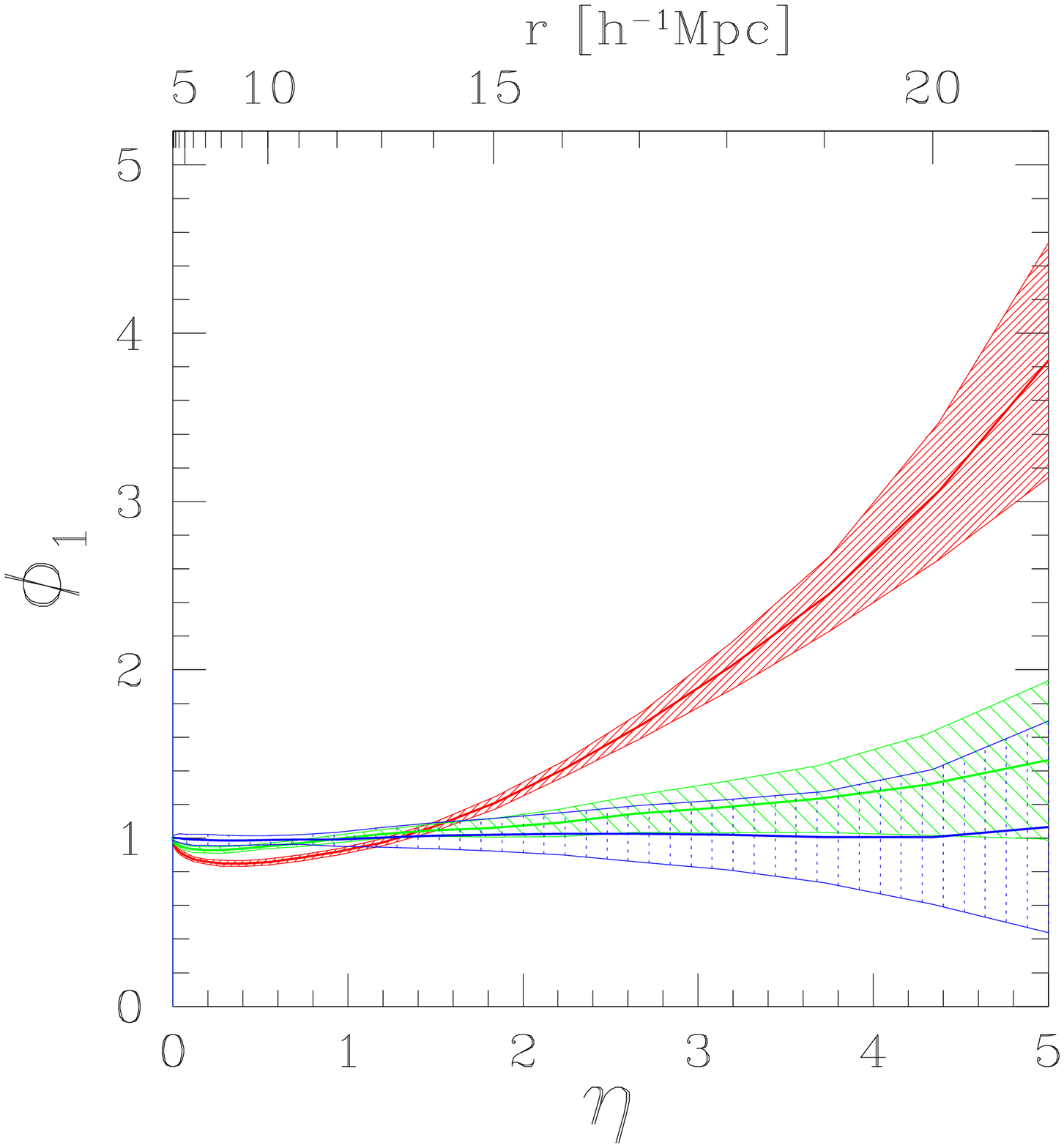}\end{minipage}
\epsfxsize=7.4cm
\begin{minipage}{\epsfxsize}\epsffile{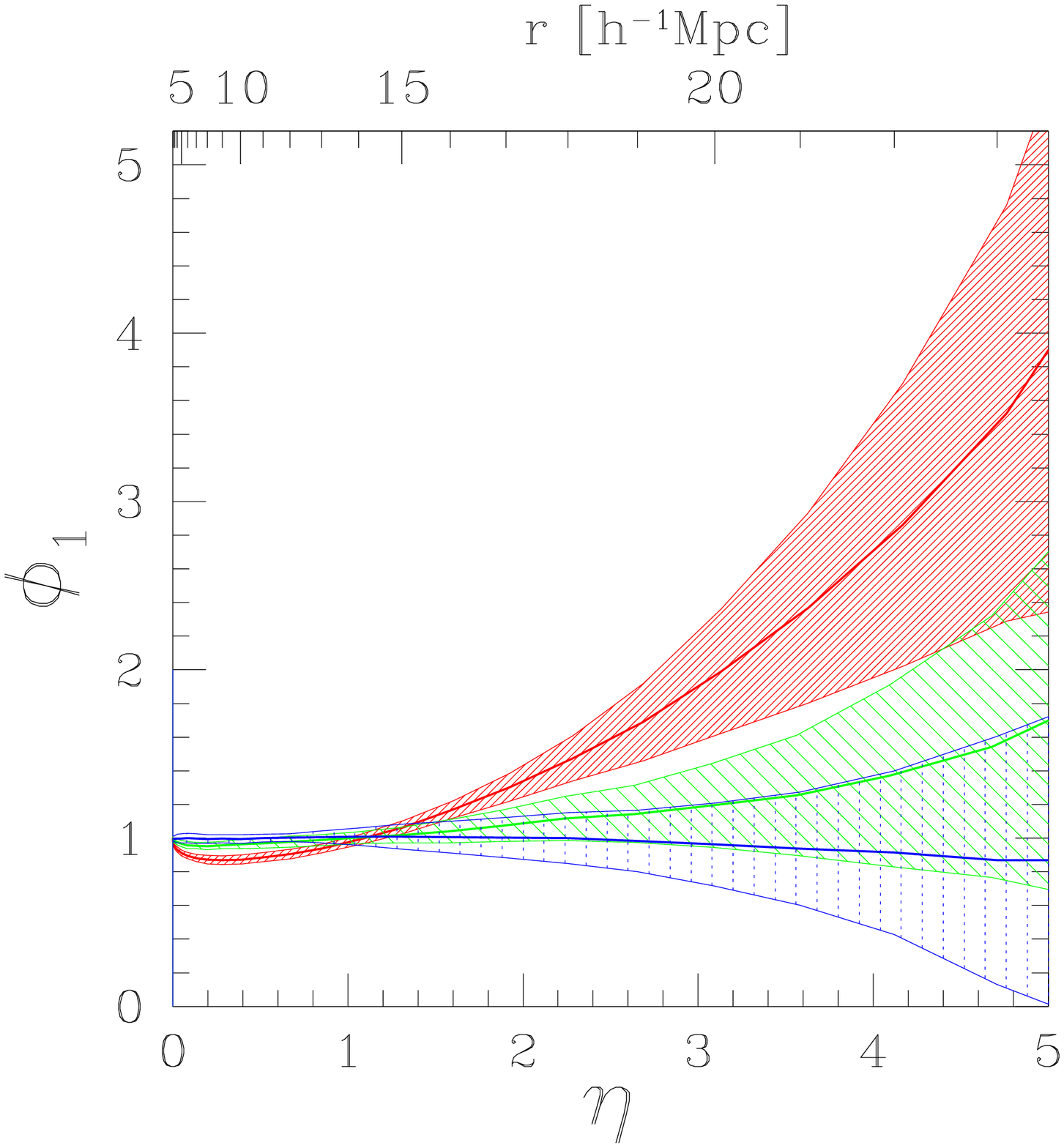}\end{minipage}
\epsfxsize=7.4cm
\begin{minipage}{\epsfxsize}\epsffile{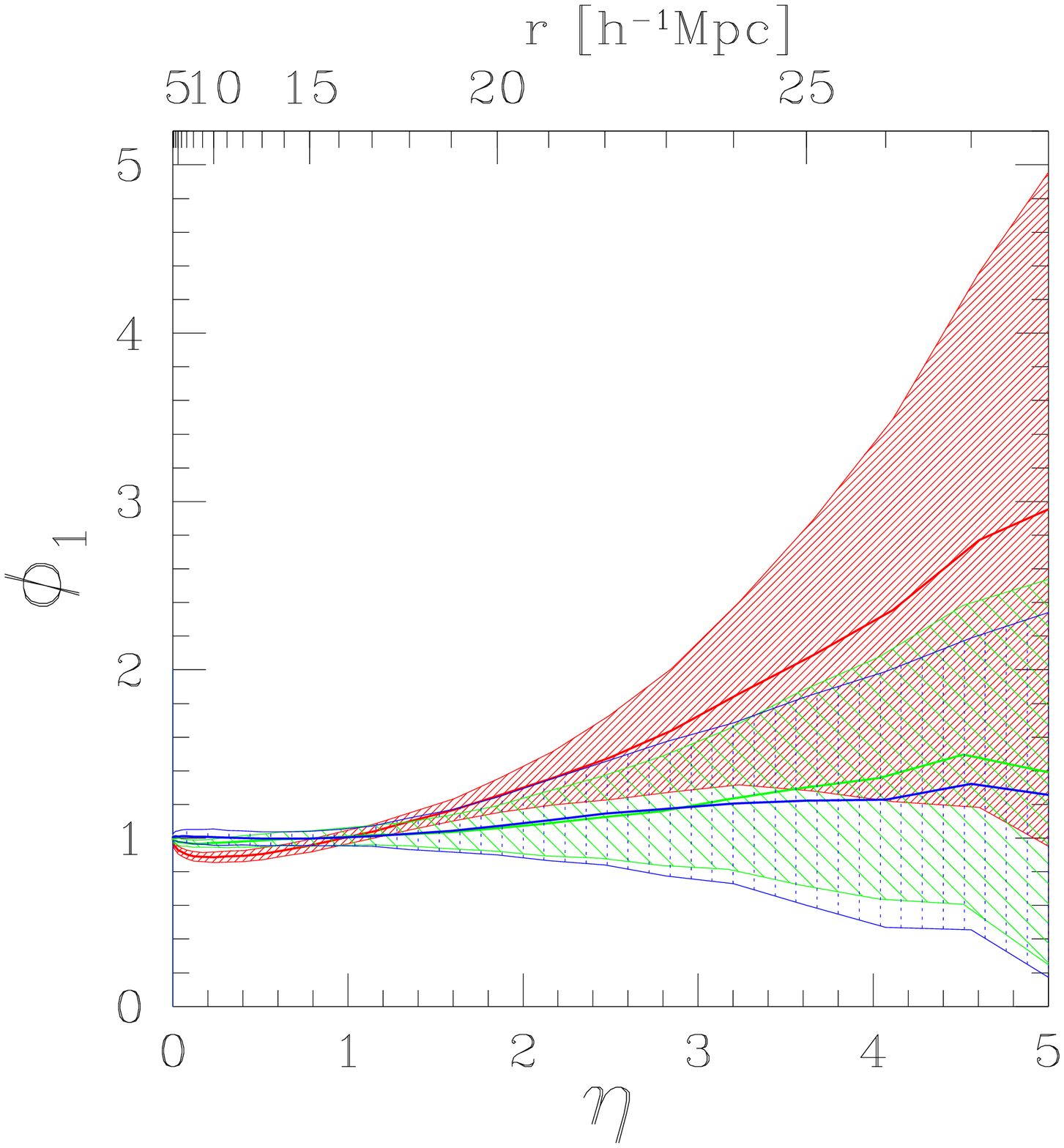}\end{minipage}
\end{center}
\caption{\label{fig:dilute}
The surface functional $\phi_1(\cA_N(r))$ of a volume limited sample
with 100\hMpc\ depth for randomly drawn subsamples with 70\% (left
figure), 50\% (middle figure), and 30\% (right figure) galaxies. Dark
shaded areas: southern part; medium shaded: northern part; dotted:
Poisson process with the same number density.}
\end{figure}

Before listing these tests, we address the important issue of `sparse
sampling'. This issue is especially relevant for the interpretation of
our tests, since all these selected samples incorporate less galaxies
than the volume limited sample with 100\hMpc\ depth. In
Figure~\ref{fig:dilute} we show how sparse sampling affects the
surface functionals $\phi_1(\cA_N(r))$ of the volume limited samples
with 100\hMpc\ by only taking a fraction of the galaxies into account
(the other functionals behave similarly). By reducing the number of
galaxies the error increases and the mean values tend towards the
values for a Poisson process. 

\begin{figure}
 \begin{center} 
 \epsfxsize=7.4cm
 \begin{minipage}{\epsfxsize}\epsffile{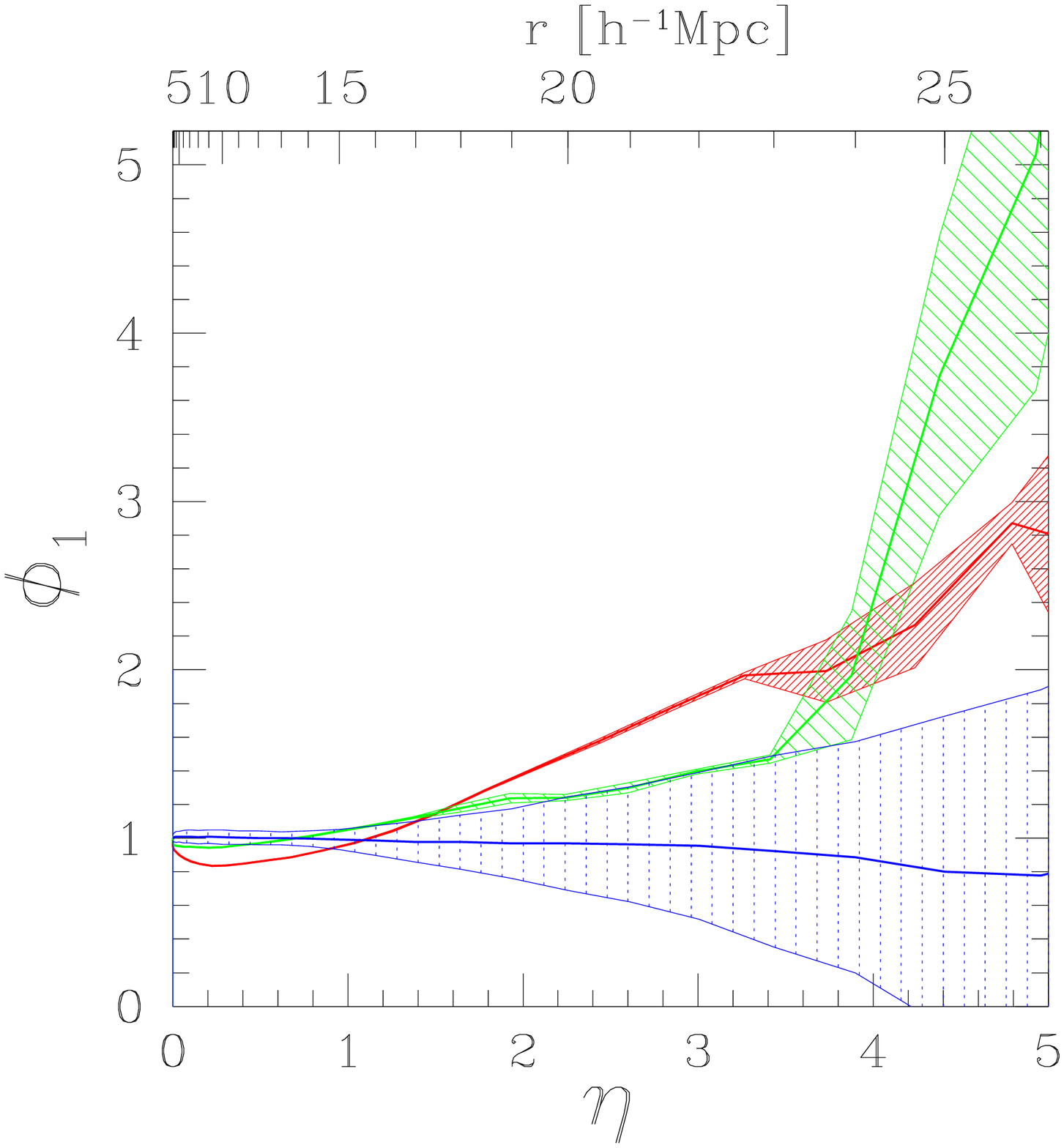}\end{minipage} 
 \epsfxsize=7.4cm
 \begin{minipage}{\epsfxsize}\epsffile{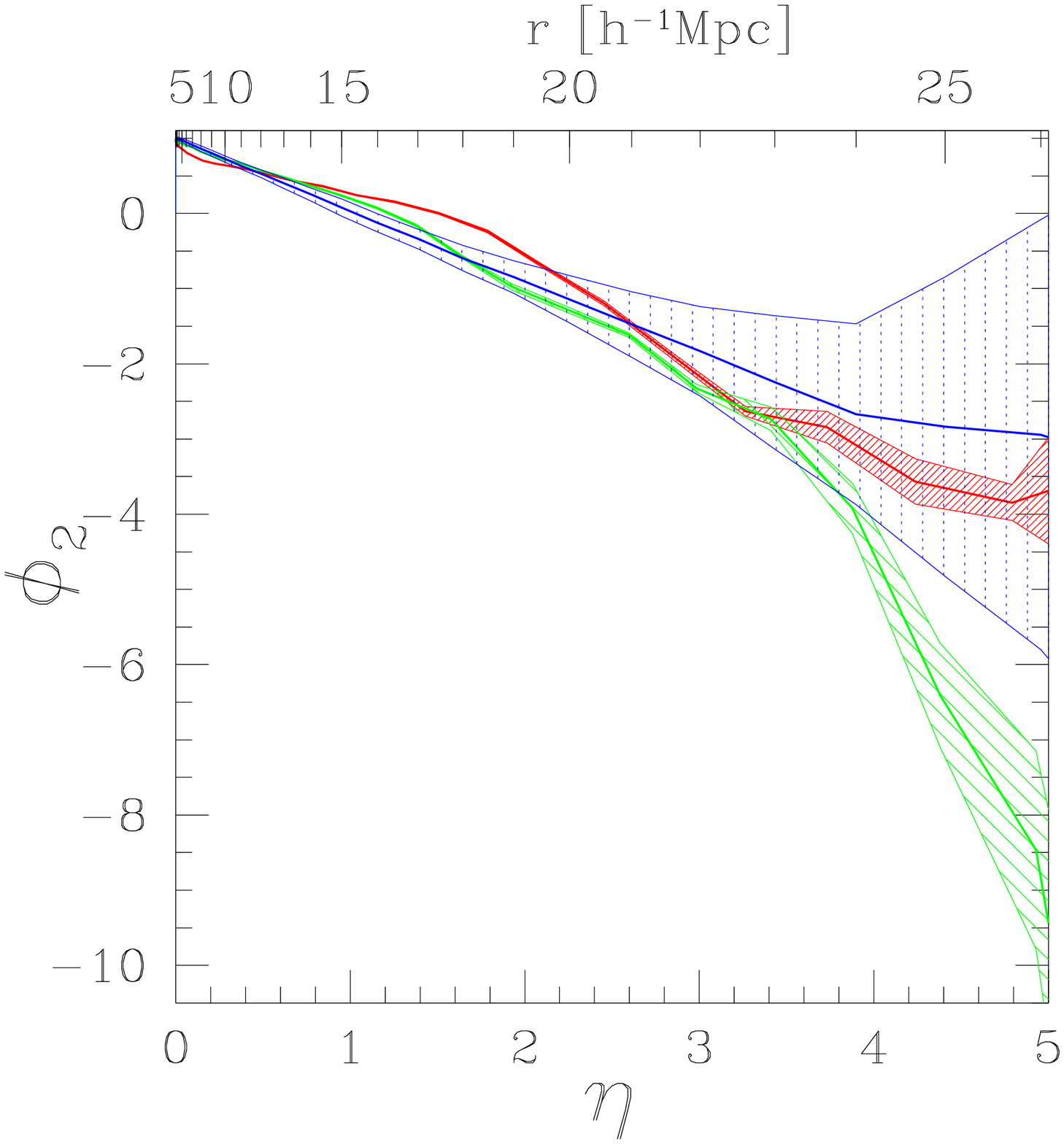}\end{minipage} 
 \epsfxsize=7.4cm
 \begin{minipage}{\epsfxsize}\epsffile{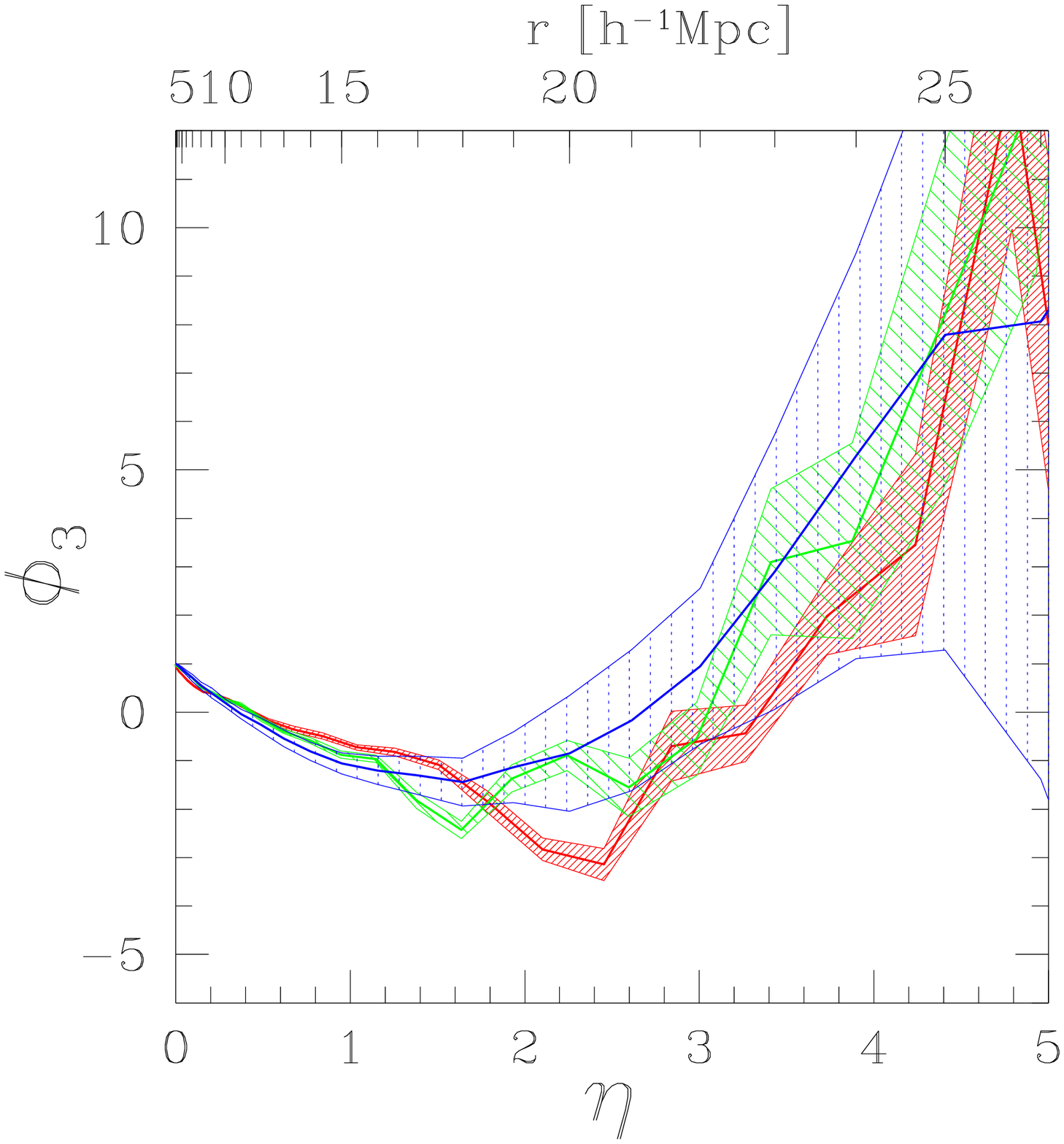}\end{minipage} 
 \end{center}
\caption{\label{fig:minjy20v10}
Minkowski functionals $\phi_\mu$ of a volume limited sample with
$100\hMpc$ depth; now the limiting flux is 2.0~Jy. The shaded areas
are the $1\sigma$ errors determined from randomizing the redshifts.}
\end{figure}

We calculated the Minkowski functionals of a volume limited sample
with 100\hMpc\ depth but now with limiting flux equal to 2.0~Jy, with
129 galaxies in the north and 141 galaxies in the south. Although the
noise increases for larger radii, since fewer galaxies enter, the
above mentioned features and the differences between northern and
southern parts are still detectable (Figure~\ref{fig:minjy20v10}).

\begin{figure}
 \begin{center} 
 \epsfxsize=7.4cm
 \begin{minipage}{\epsfxsize}\epsffile{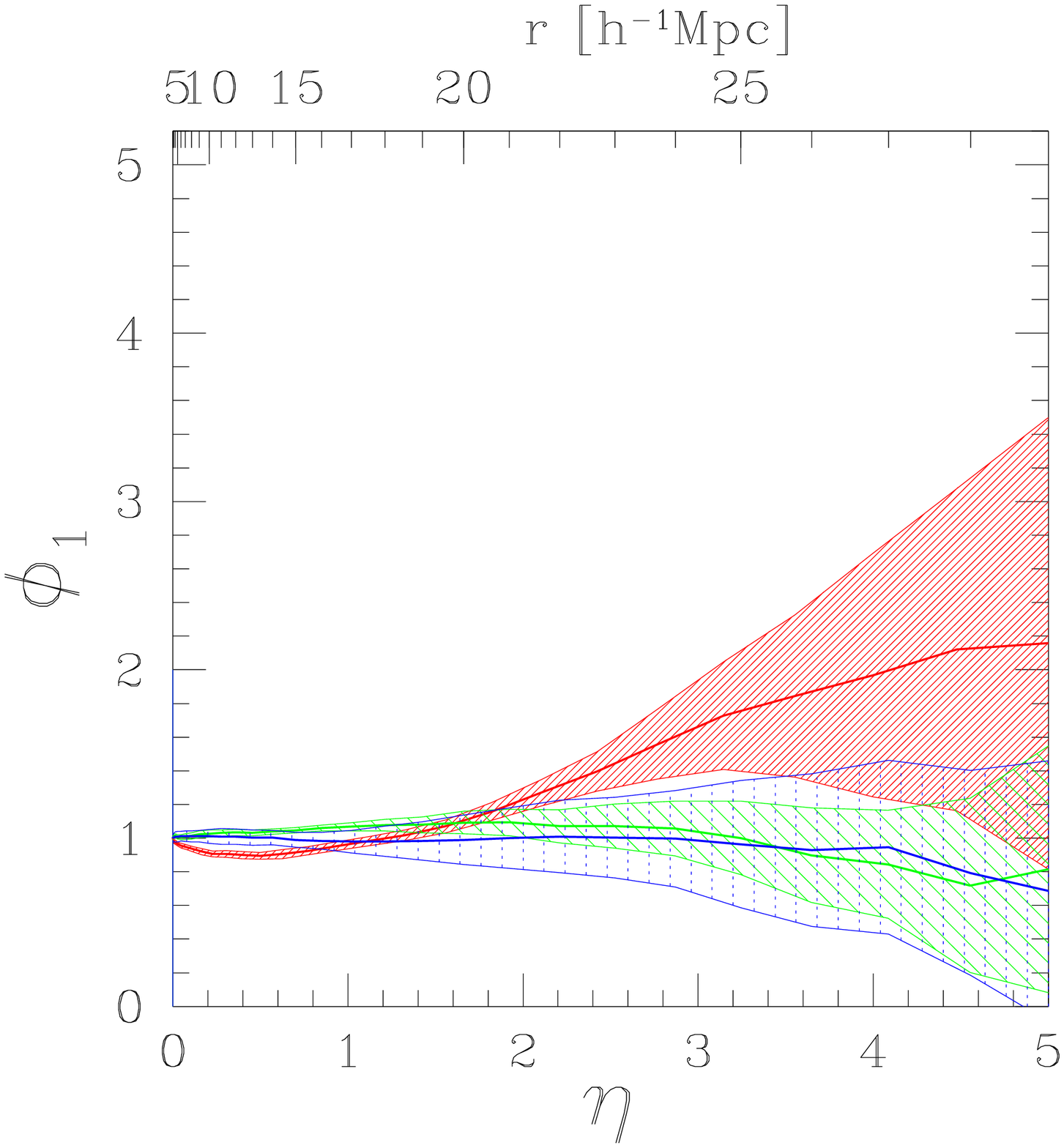}\end{minipage} 
 \epsfxsize=7.4cm
 \begin{minipage}{\epsfxsize}\epsffile{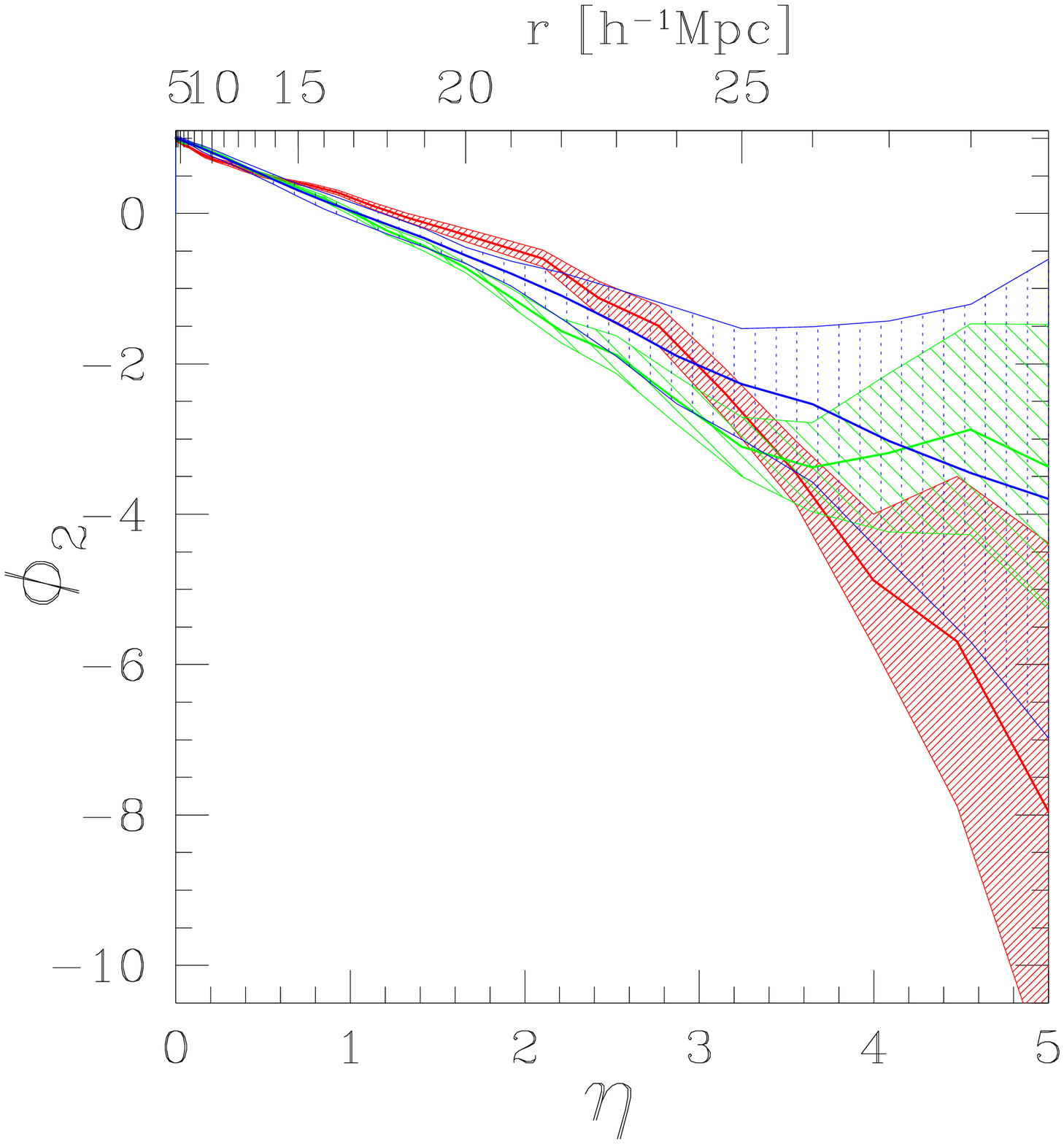}\end{minipage} 
 \epsfxsize=7.4cm
 \begin{minipage}{\epsfxsize}\epsffile{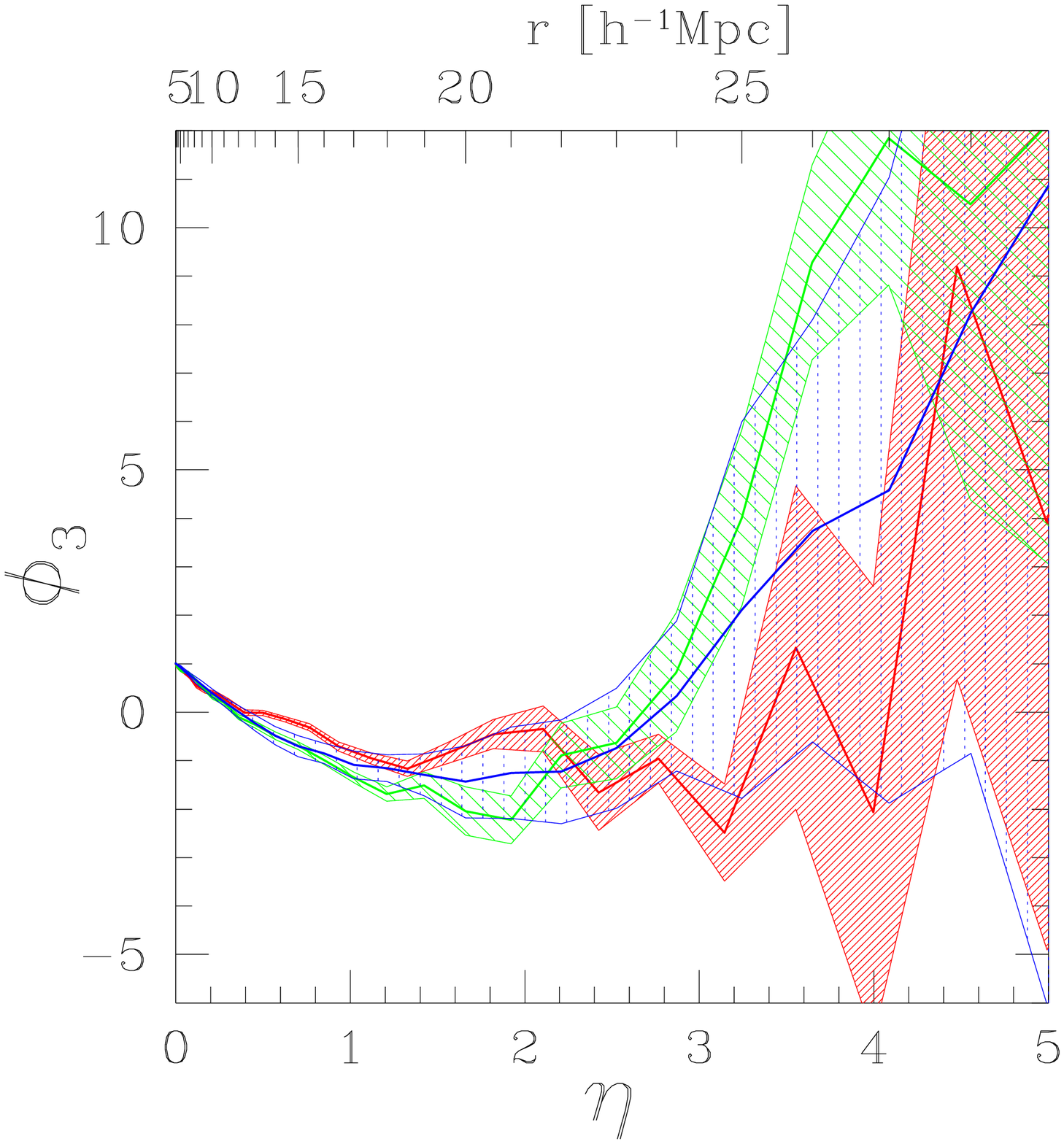}\end{minipage} 
 \end{center}
\caption{\label{fig:minjyhot}
Minkowski functionals $\phi_\mu$ of a volume limited sample with
$100\hMpc$ depth with only ``hot'' galaxies included (limiting flux
1.2 Jy). The shaded areas are the $1\sigma$ errors as in
Figure~\ref{fig:minjyv10}.}
\end{figure}

\begin{figure}
 \begin{center} 
 \epsfxsize=7.4cm
 \begin{minipage}{\epsfxsize}\epsffile{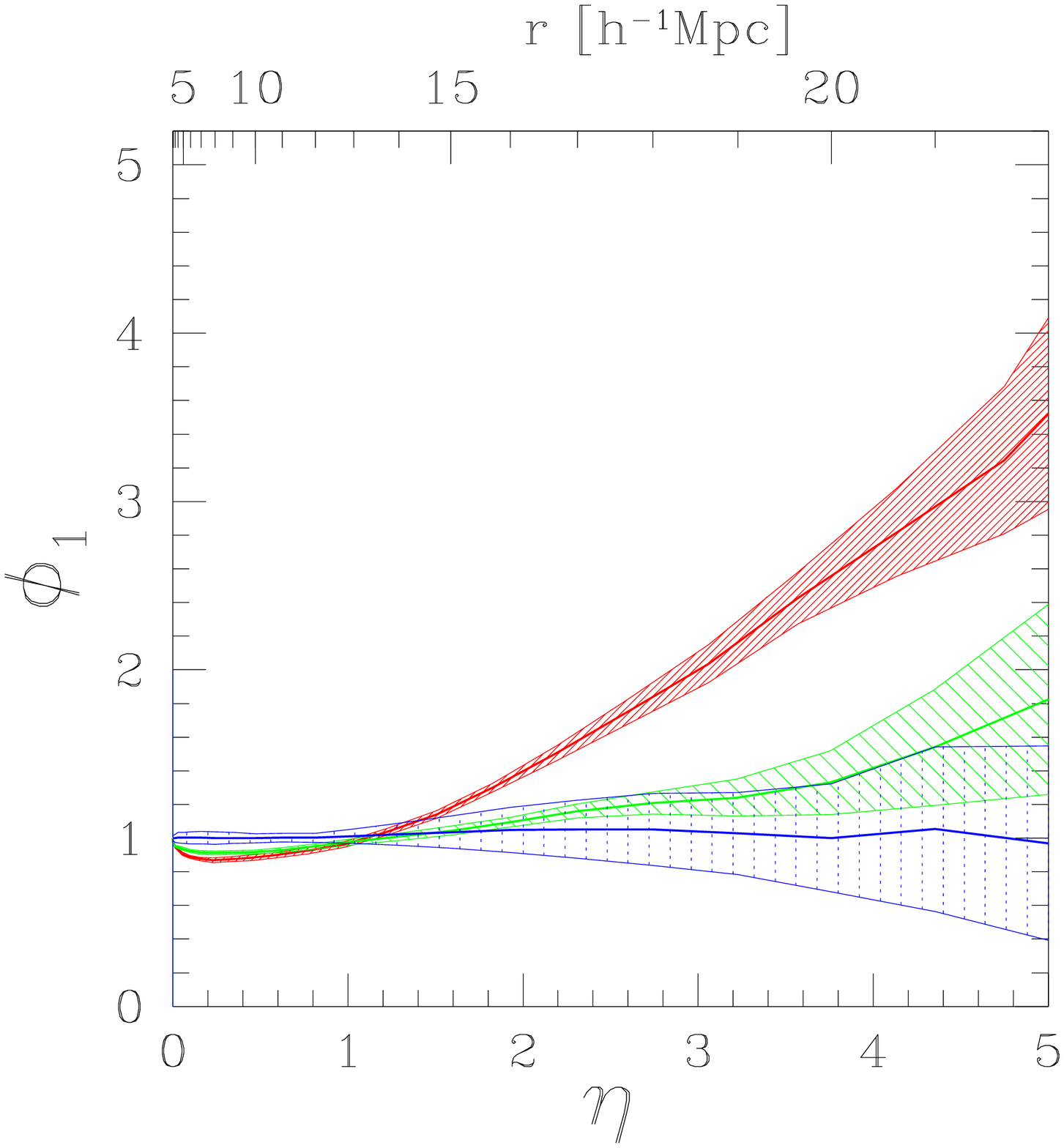}\end{minipage} 
 \epsfxsize=7.4cm
 \begin{minipage}{\epsfxsize}\epsffile{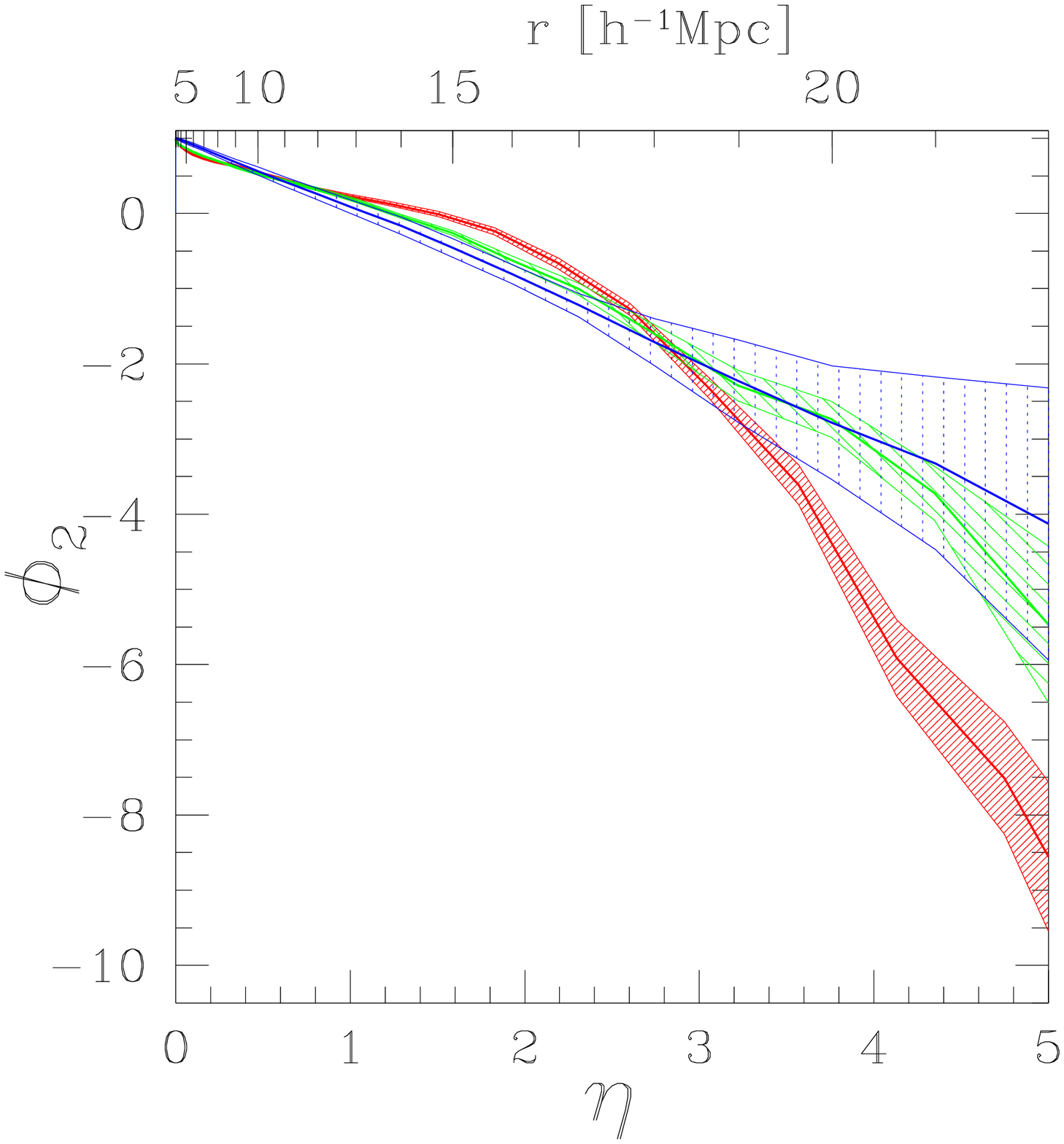}\end{minipage} 
 \epsfxsize=7.4cm
 \begin{minipage}{\epsfxsize}\epsffile{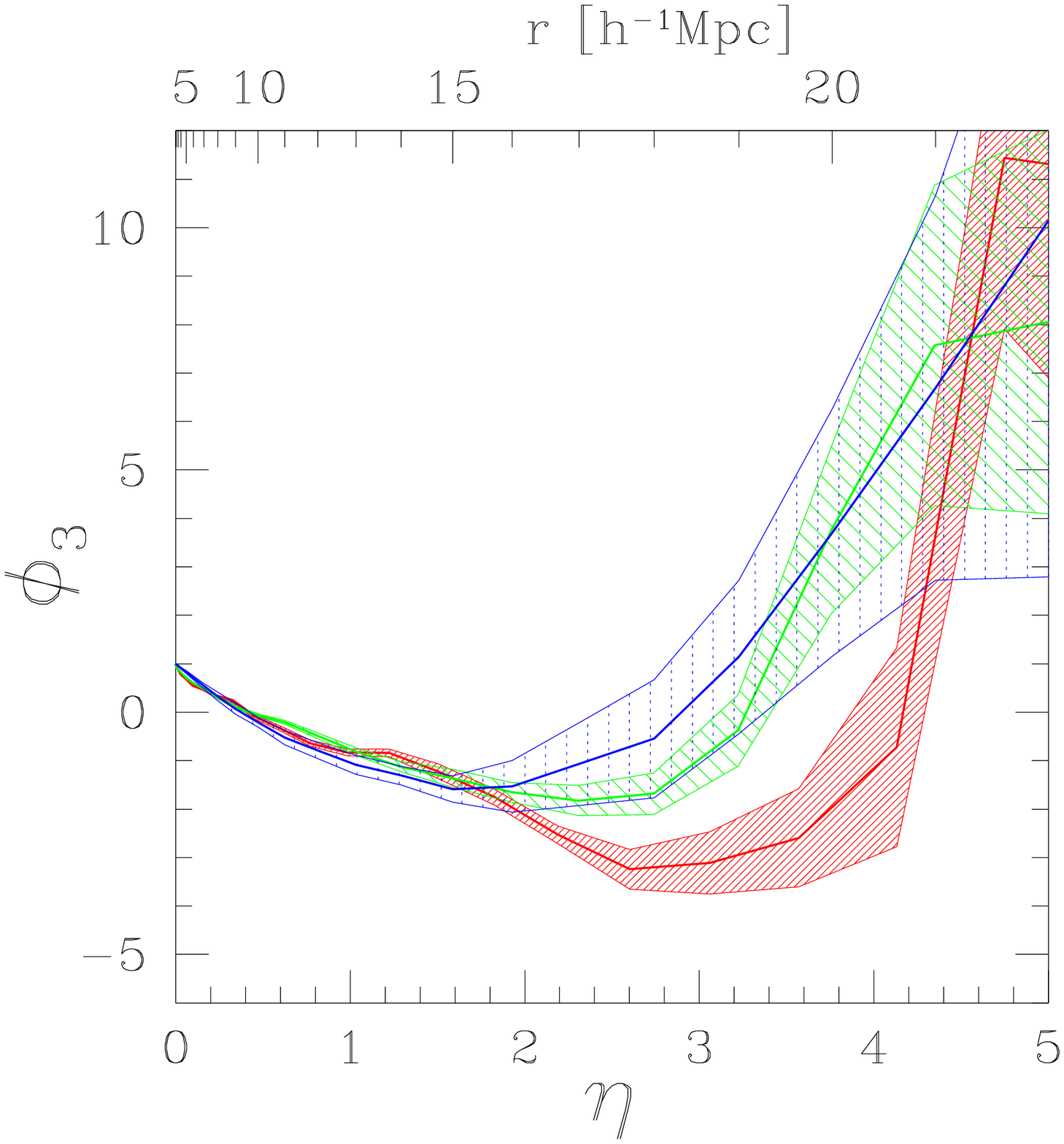}\end{minipage} 
 \end{center}
\caption{\label{fig:minjymed}
Minkowski functionals $\phi_\mu$ of a volume limited sample with
$100\hMpc$ depth with only ``warm'' galaxies included (limiting flux
1.2 Jy). The shaded areas are the $1\sigma$ errors as in
Figure~\ref{fig:minjyv10}.}
\end{figure}

We selected ``hot'' galaxies, with a flux ratio $f_{100}/f_{60}
\leq 1.5$, ``warm'' ones with $1.5 \geq f_{100}/f_{60} \leq 3$ and ``cold'' 
galaxies with $f_{100}/f_{60} \geq 3$ from the 1.2~Jy catalogue;
$f_{100}$ and $f_{60}$ denote the flux at $100\mu$ and $60\mu$
respectively. We calculated the Minkowski functionals of volume
limited samples with 100\hMpc\ taken from the ``hot'' (106 in the
north and 116 in the south) and ``warm'' (239 in the north and 227 in
the south) galaxies (only 7 (north) and 15 (south) galaxies are
``cold''). Again the error increases, but the features and the
differences remain visible (compare Figure~\ref{fig:minjyhot} and
{}\ref{fig:minjymed}). With more refined but essentially similar
criteria for distinguishing ``warm'' and ``cool'' galaxies
{}\scite{mann:warmcool} find only a small dependence of clustering on
the temperature in the QDOT.


\newpage

\section{Where do fluctuations originate?}
\label{sect:origin}
\begin{figure}
 \begin{center} 
 \epsfxsize=8cm
 \begin{minipage}{\epsfxsize}\epsffile{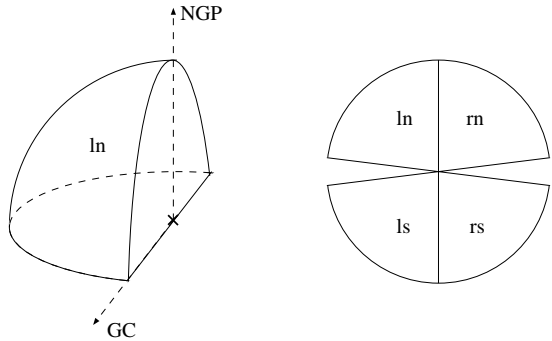}\end{minipage} 
 \end{center}
\caption{\label{fig:cut}
Sketch of the sample cut along the plane through our position, the
north galactic pole (NGP) and the galactic center (GC). The figure on
the right shows how we label them.}
\end{figure}

Now we try to find out where the fluctuations have their spatial
origin. To do this we cut the north an south into two parts each, as
shown in Fig.~\ref{fig:cut} and in Table~\ref{table:cut}. The four
samples approximately contain 160 galaxies.
\begin{table}
\begin{center}
\begin{tabular}{|c|c|c|}
\hline
name & lattitude range & longitude range \\
\hline
{\em rn}   & $5^\circ \le b^{\rm II} \le 90^\circ$  & 
  $0^\circ \le l^{\rm II} \le 180^\circ$   \\
{\em ln}   & $5^\circ \le b^{\rm II} \le 90^\circ$  & 
  $180^\circ \le l^{\rm II} \le 360^\circ$ \\
{\em rs}   & $-5^\circ \ge b^{\rm II} \ge -90^\circ$  & 
  $0^\circ \le l^{\rm II} \le 180^\circ$   \\
{\em ls}   & $-5^\circ \le b^{\rm II} \ge -90^\circ$  & 
  $180^\circ \le l^{\rm II} \le 360^\circ$   \\
\hline
\end{tabular}
\end{center}
\caption{\label{table:cut}
The angular ranges covered by the four samples shown in
Fig.~\ref{fig:cut}.}
\end{table}

\begin{figure}
 \begin{center} 
 \epsfxsize=7.4cm
 \begin{minipage}{\epsfxsize}\epsffile{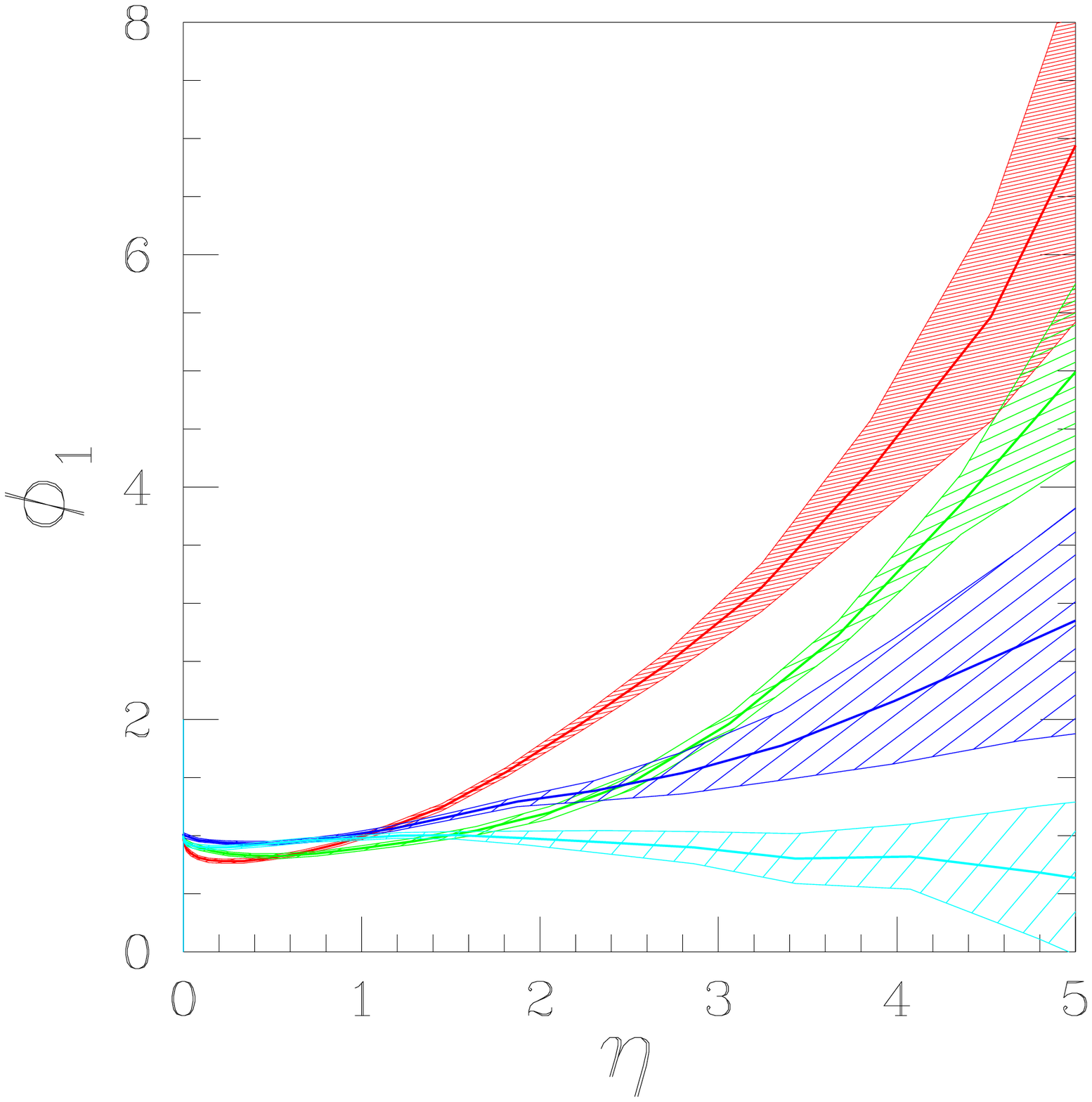}\end{minipage} 
 \epsfxsize=7.4cm
 \begin{minipage}{\epsfxsize}\epsffile{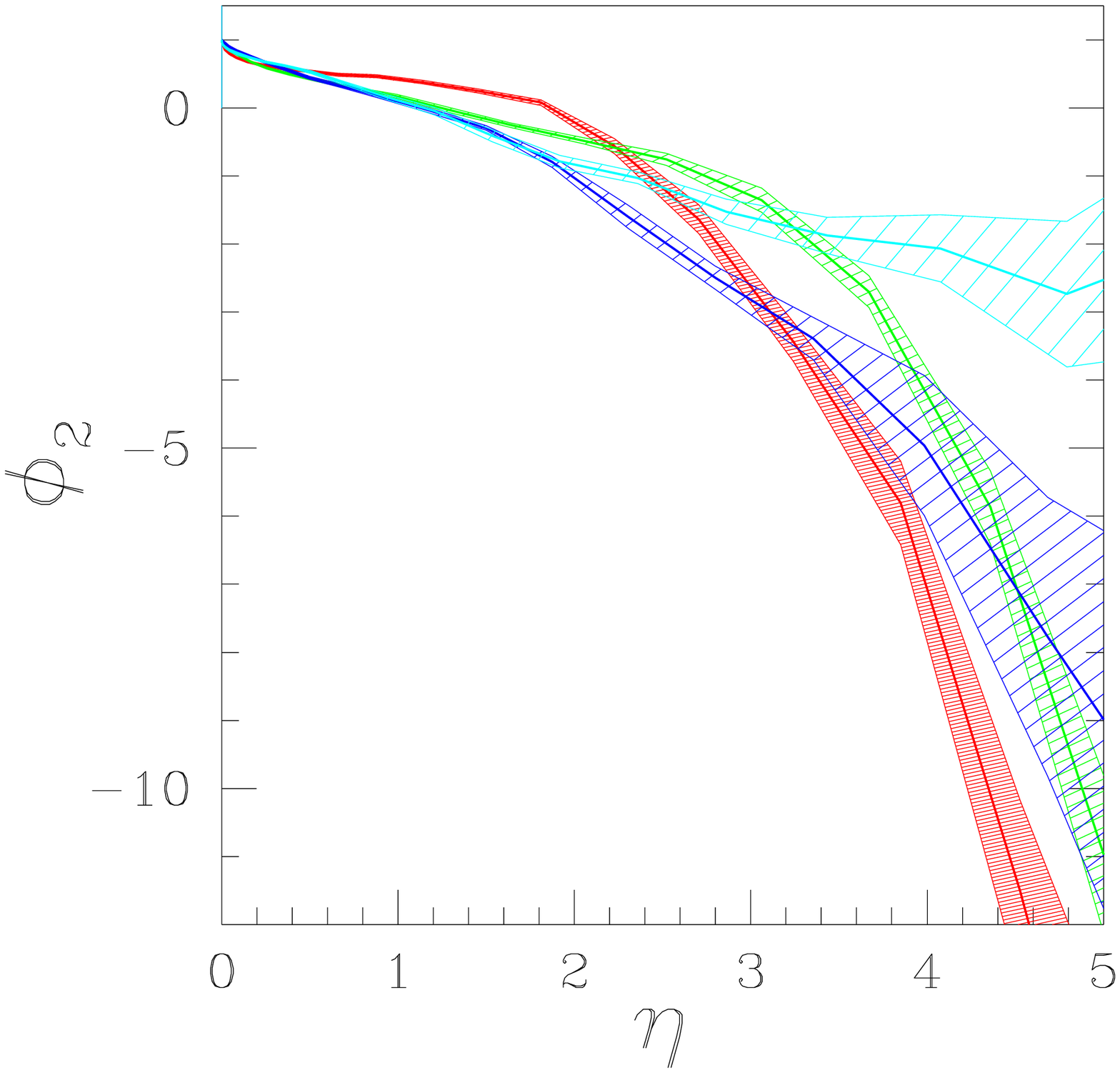}\end{minipage} 
 \epsfxsize=7.4cm
 \begin{minipage}{\epsfxsize}\epsffile{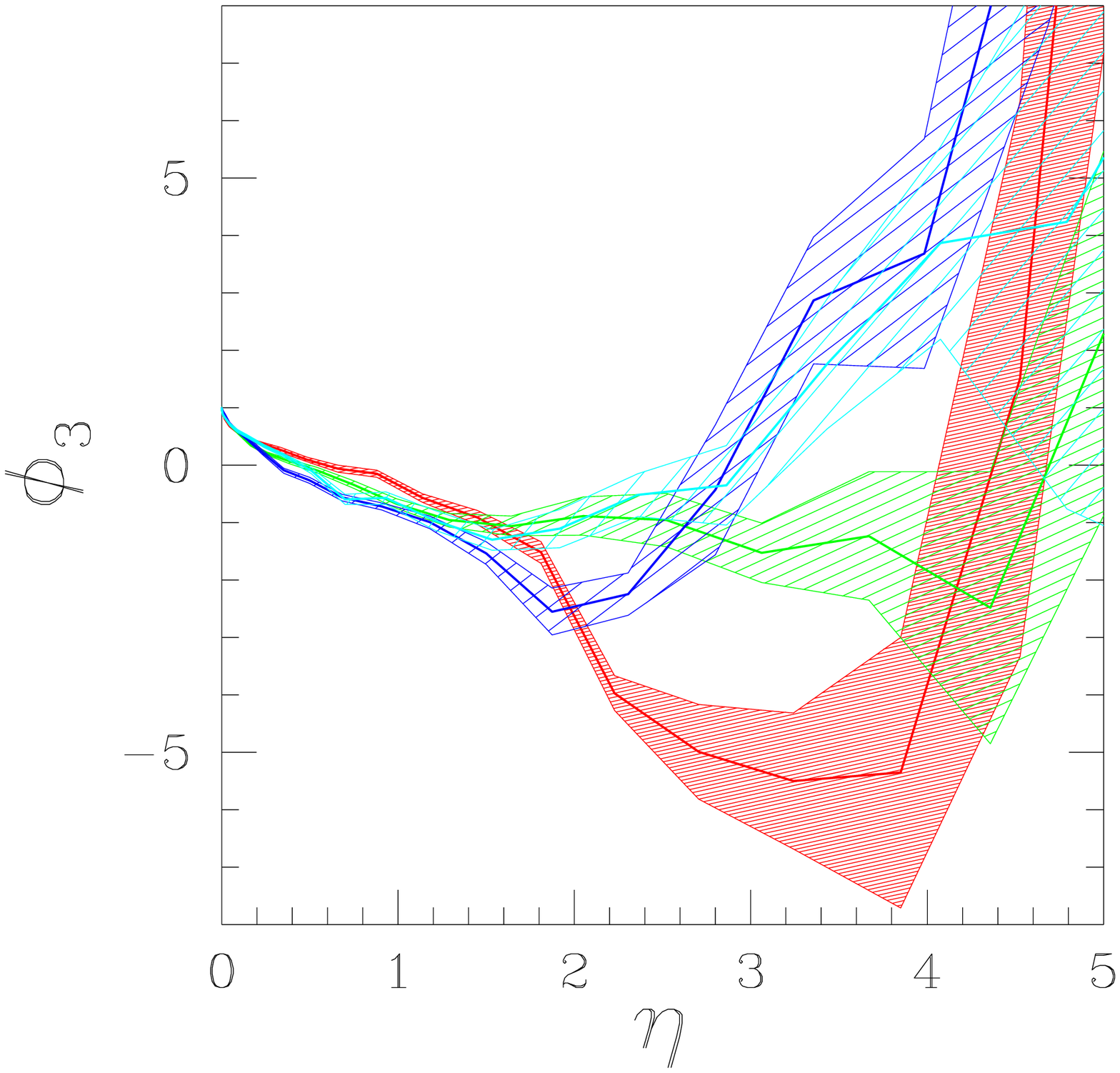}\end{minipage} 
 \end{center}
\caption{\label{fig:minjyhalf}
Minkowski functionals $\phi_\mu$ of a volume limited sample with
$100\hMpc$ depth; cut into pieces as shown in
Fig.~\protect\ref{fig:cut}. The shading of the $1\sigma$ areas
(determined from subsampling 90\% of the galaxies) is getting darker
in the following ordering of the samples {\em ln, rn, rs, ls}.}
\end{figure}

Fig.~\ref{fig:minjyhalf} shows that all four parts differ from each
other. Both southern parts cluster more strongly than the northern
parts, consistent with Fig.~\ref{fig:minjyv10}. The strongest
clustering is seen in the sample {\em ls}, the weakest clustering is
in the sample {\em ln}. The Perseus--Pegasus--Pisces Supercluster lies
in the sample {\em rs}.

\section{Comparison with the CfA1}
\label{sect:jy-cfa}

An objection against results based on IRAS selected samples is that we
mainly look at infrared active, spiral galaxies. Therefore we compared
samples from the CfA1 galaxy catalogue with the samples from the
1.2~Jy catalogue.
We consider volume--limited samples extracted from CfA1
{}\cite{huchra:cfa1}, whose members have apparent magnitudes less than
14.5. The limits of the samples considered are galactic latitude
$b>40\deg$ and declination $\delta>0\deg$, the volume limitation is
performed with 100\hMpc\ depth leaving 99 galaxies within the sample
geometry. The volume limited sample of the 1.2~Jy with 100\hMpc\ has
115 galaxies within the CfA1 window. To compare both samples we draw
random subsamples from the 1.2~Jy to reproduce the number density in
the CfA1 sample.

\begin{figure}
 \begin{center} 
 \epsfxsize=7.4cm
 \begin{minipage}{\epsfxsize}\epsffile{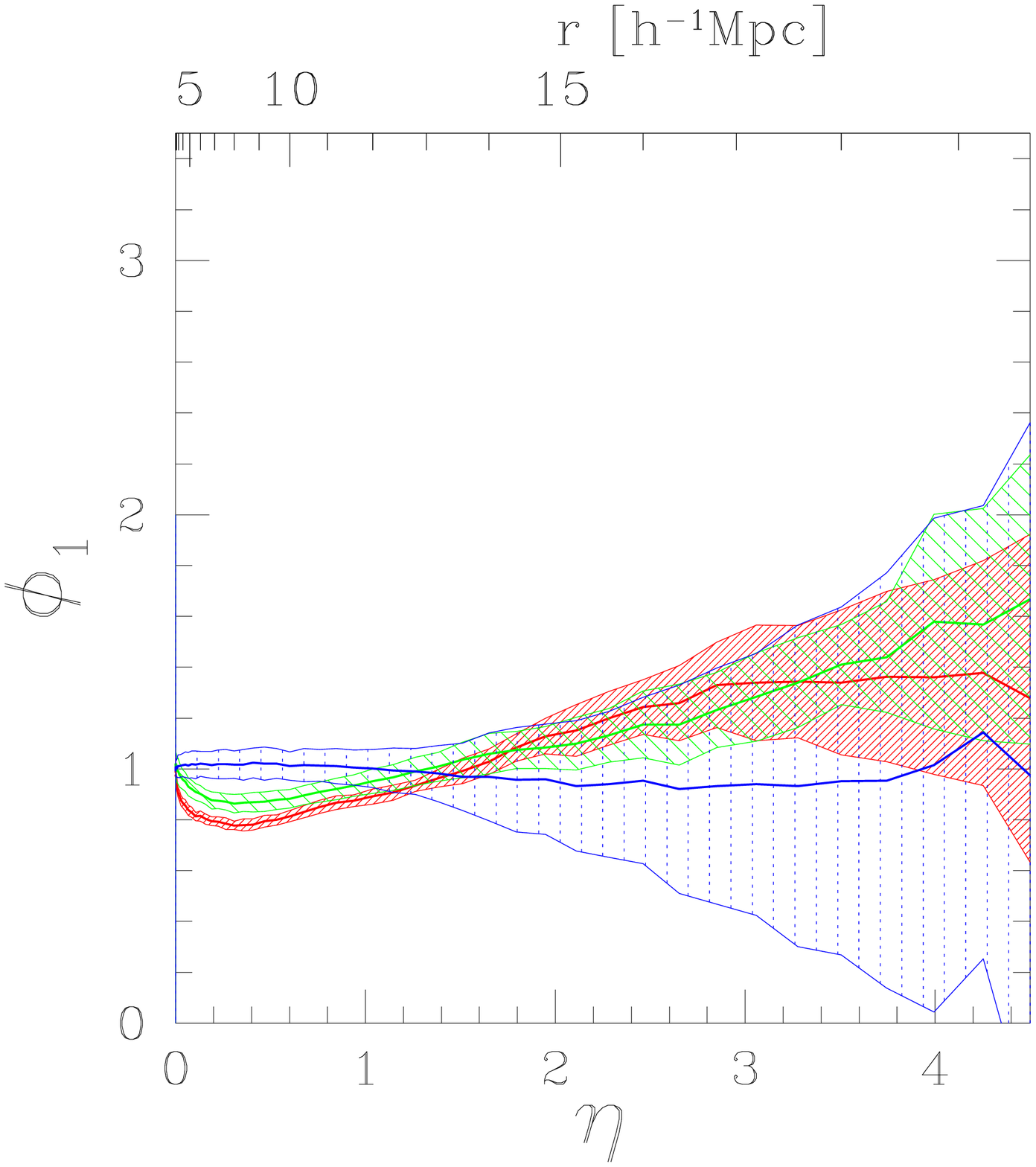}\end{minipage}
 \epsfxsize=7.4cm
 \begin{minipage}{\epsfxsize}\epsffile{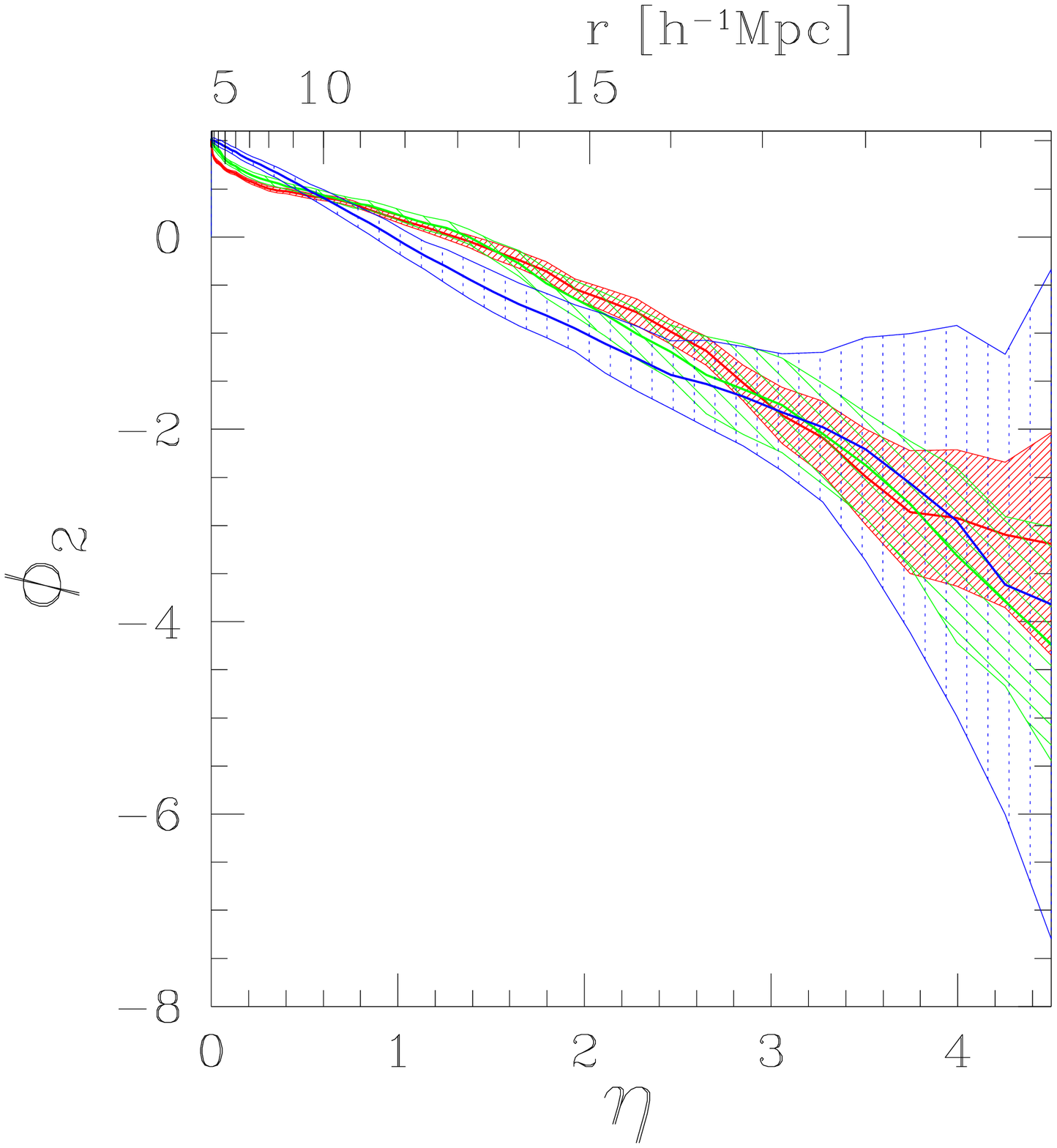}\end{minipage} 
 \epsfxsize=7.4cm
 \begin{minipage}{\epsfxsize}\epsffile{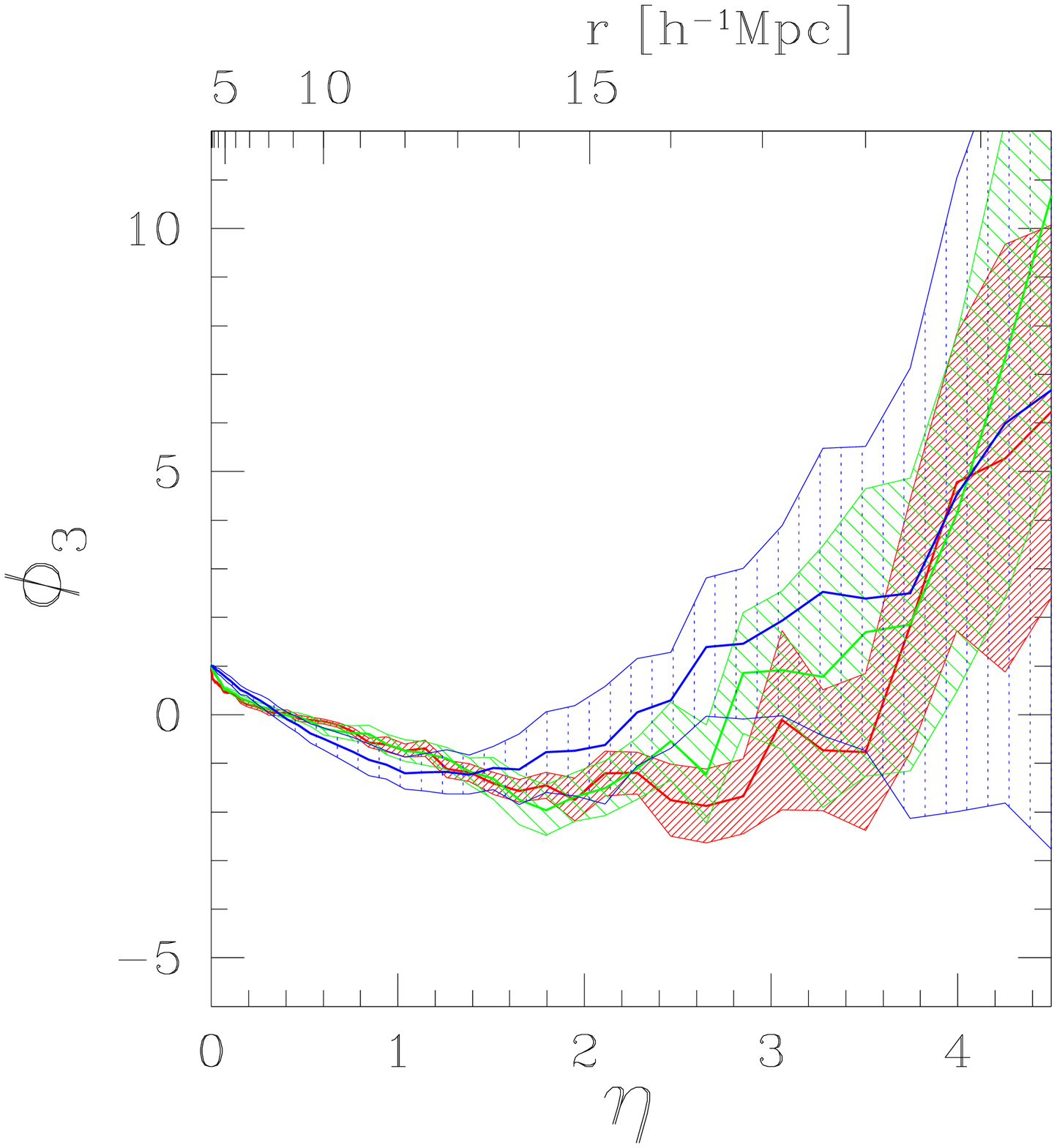}\end{minipage} 
 \end{center}
\caption{\label{fig:jy-cfa10}
Minkowski functionals $\phi_\mu$ of a volume limited sample with
$100\hMpc$ depth of CfA1 (dark shaded) compared to the Minkowski
functionals of 1.2~Jy (medium shaded) and a Poisson process (light
shaded).The shaded areas are the $1\sigma$ errors determined from
random subsampling 90\% of the galaxies of the CfA1 and 77\% of the
1.2~Jy.}
\end{figure}

The 1.2~Jy galaxy catalogue mainly consists of spiral galaxies, and is
therefore undersampling the cluster cores. In complete agreement with
this, we see in Figure~\ref{fig:jy-cfa10} less clustering in the
1.2~Jy on scales up to 10\hMpc, deduced from the lowered surface
functional $\phi_1$. On large scales both samples are compatible, no
systematic difference is seen between the IRAS and the optical sample.
On small scales both samples differ from the Poisson results but are
compatible with Poisson on large scales. 

\begin{figure}
 \begin{center} 
 \epsfxsize=7.4cm
 \begin{minipage}{\epsfxsize}\epsffile{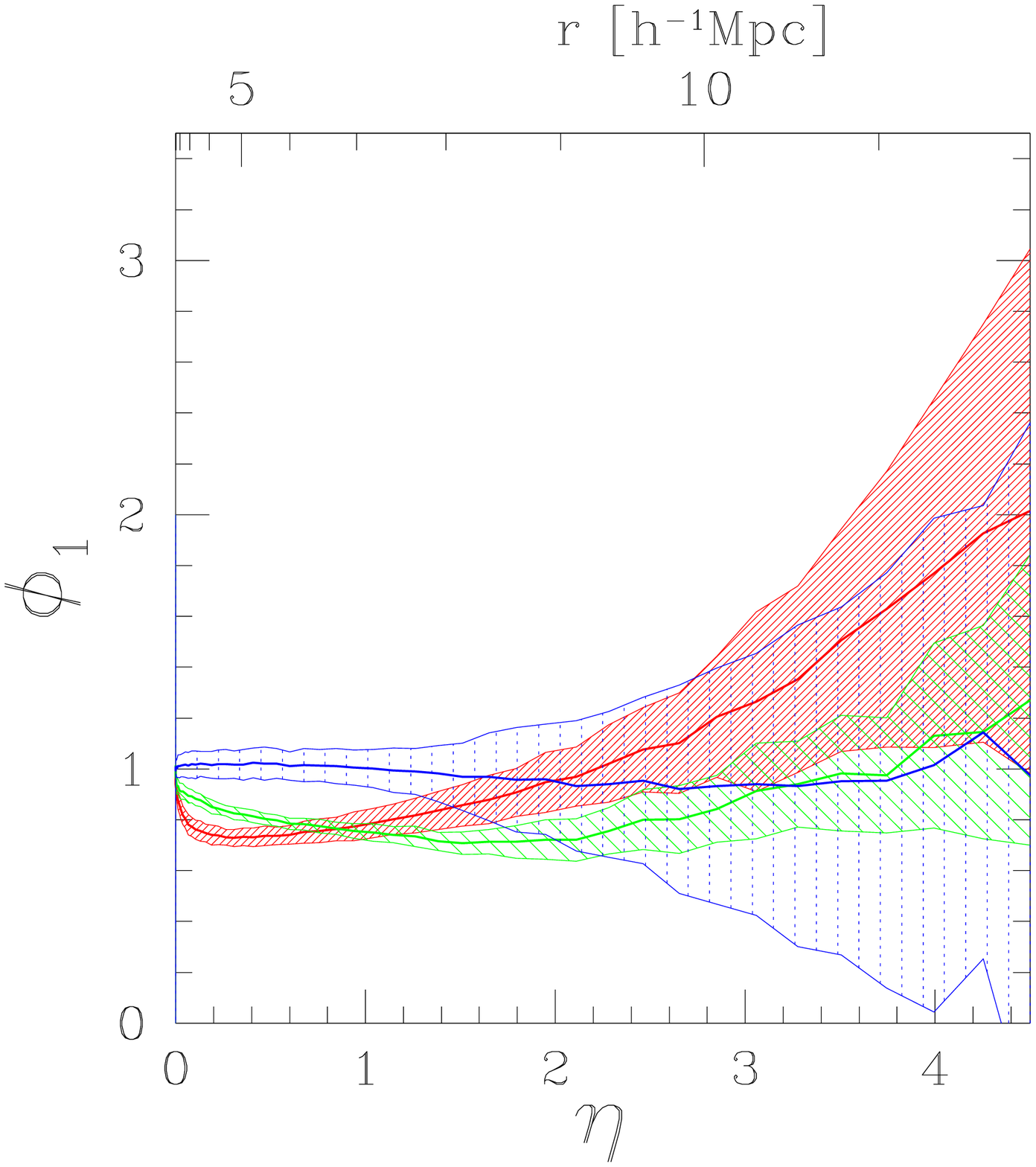}\end{minipage}
 \epsfxsize=7.4cm
 \begin{minipage}{\epsfxsize}\epsffile{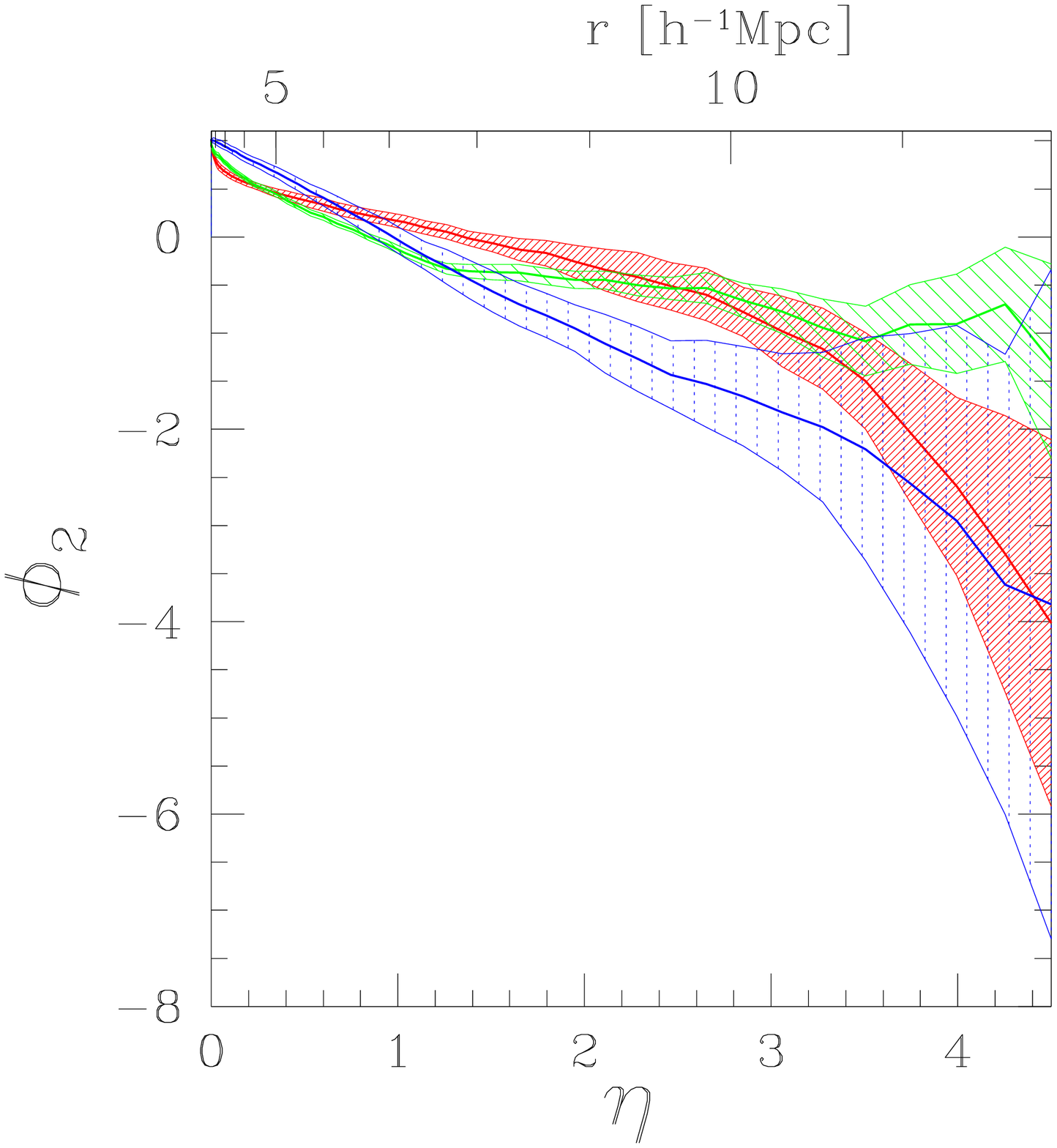}\end{minipage} 
 \epsfxsize=7.4cm
 \begin{minipage}{\epsfxsize}\epsffile{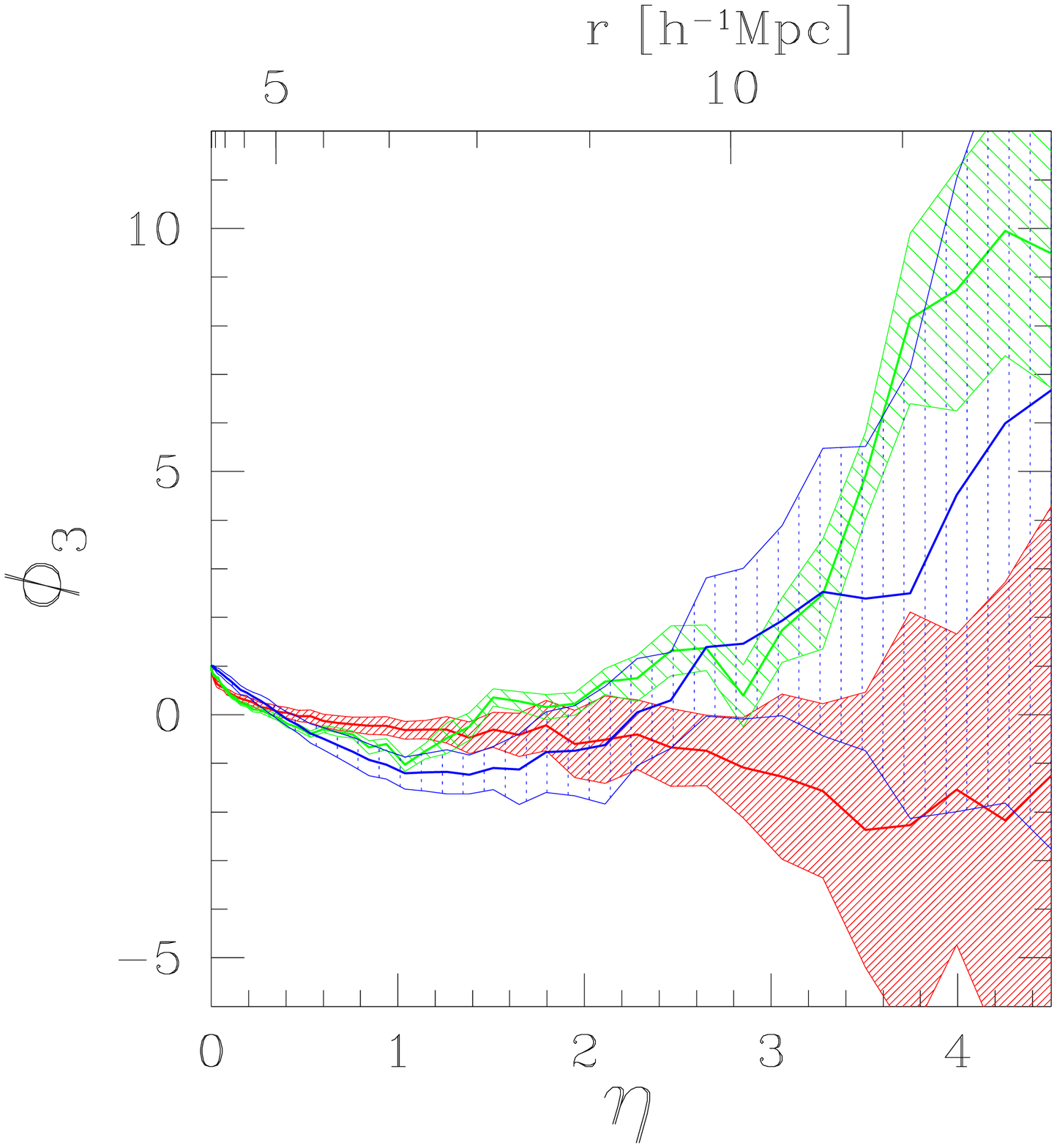}\end{minipage} 
 \end{center}
\caption{\label{fig:jy-cfa6}
Minkowski functionals $\phi_\mu$ of a volume limited sample with
$60\hMpc$ depth of CfA1 (dark shaded) compared to the Minkowski
functionals of 1.2~Jy (medium shaded) and a Poisson process (light
shaded).The shaded areas are the $1\sigma$ errors determined from
random subsampling 42\% of the galaxies of the CfA1 and 90\% of the
1.2~Jy.}
\end{figure}

Since we are severely affected by sparse sampling (see
Section~\ref{sect:selection}) we checked the above detected
concordance of the morphological properties of optical and IRAS
selected galaxies on large scales in looking at samples with 60\hMpc\
depth. Now the CfA1 sample includes 215 galaxies and the 1.2~Jy
sample 115. In Figure~\ref{fig:jy-cfa6} a tendency towards stronger
clustering of the optically selected galaxies is seen on all scales. One
has to bear in mind, that we compared two samples with only 22\% of
the volume of the samples with 100\hMpc\ depth. Hence it is not clear
whether this is an effect only seen in our ``local surrounding''.
To do a reliable comparison of the morphology of optical and infrared
selected galaxy distributions we have to wait for larger optical and
infrared redshift surveys.

\section{Discussion and conclusions}
\label{sect:conclusion}
We analyzed morphological characteristics of the galaxy distribution
given by the IRAS 1.2~Jy catalogue. The two subsamples (north and
south) of this catalogue were studied individually with
Minkowski functionals.

Since the IRAS data have been obtained from a single instrument with
uniform calibration, the two subsamples, which contain about the same
number of galaxies, can be compared reliably.

We have reported in detail our error estimates and have discussed
tests on selection effects to assess the significance of our
results. For reference purposes we used typical realizations of a
stationary Poisson point process.

Our results can be summarized as follows: The values of the Minkowski
functionals for the southern part differ significantly from those for
the northern part in the volume limited subsample with 100\hMpc\
depth. This difference does not disappear on scales of 200\hMpc\ as
shown in {}\scite{kerscher:fluctuations}.

Similar anisotropies in the angular distribution of IRAS galaxies
around the northern and southern galactic poles have already been
reported by {}\scite{rowan:sourcecount}. However, the majority of
previous IRAS catalogue studies have focused attention on the complete
sample without addressing the distinction of its constituent parts.

There is no reason to assume a distinguished position of the Milky Way
galaxy; we therefore conclude that fluctuations in the global
morphological characteristics of the IRAS sample extend over
length scales of at least 100\hMpc. These fluctuations may originate
from dynamical correlations in the matter distribution which arise
during the cosmic evolution.

During the last few years, {}\scite{coleman:fractal}, see also
{}\scite{labini:frequently} in this volume, have advanced an
interpretation of galaxy catalogue data in terms of a fractal support
of the galaxy distribution. By its nature, a pure fractal indeed
exhibits fluctuations in $N$--point distributions on all scales. Our
results neither support nor contradict this interpretation, since the
Minkowski functionals, as employed in the present paper, are global
measures and are not designed to discriminate local structures of
spatial patterns.

Fluctuations occuring on scales up to 100\hMpc, at least, imply that
cosmological simulations which {\em enforce} homogeneity on the scale
of a few hundreds of \hMpc\ and suppress fluctuations on larger scales
by using periodic boundary conditions cannot reproduce the
large--scale fluctuations indicated by the present analysis of the
1.2~Jy catalogue. This assertion is confirmed by our comparison with
IRAS mock catalogues drawn from simulations of 256\hMpc\ box--length
\cite{kerscher:fluctuations}.

It remains an open question, whether it is possible to explain these
observed fluctuations in the clustering properties within large
realizations of the standard model and with COBE normalized Gaussian
initial density fields.

\section*{Acknowledgements}
MK and TB acknowledge support from the {\em Sonderforschungsbereich
SFB 375 f\"ur Astroteilchenphysik der Deutschen
Forschungsgemeinschaft} and from Acci\'on Integrada Hispano--Alemana
HA-188A (MEC).
We thank Roberto Trassarti--Battistoni for valuable comments.


\providecommand{\bysame}{\leavevmode\hbox to3em{\hrulefill}\thinspace}

\end{document}